\documentclass[12pt, review]{elsarticle}
\usepackage[a4paper]{geometry}
\usepackage{graphicx}
\usepackage{helvet}
\usepackage[hidelinks]{hyperref}
\usepackage{amsmath} 
\usepackage[italic]{mathastext}
\usepackage{amssymb} 

\usepackage{ifthen}
\usepackage[per-mode=symbol]{siunitx}\usepackage[version=4]{mhchem}
\usepackage[capitalise,nameinlink]{cleveref}

\DeclareSIUnit{\EUR}{\text{\euro}}
\usepackage{url}
\usepackage{floatrow}
\usepackage{multirow} 
\usepackage{pdflscape} 
\usepackage{eurosym}
\usepackage[toc,page,titletoc]{appendix}
\usepackage{ifthen}

\usepackage{tikz}

\usepackage{framed} 
\usepackage{multicol} 

\usepackage[noprefix]{nomencl} 
\makenomenclature
\newboolean{showfigures}
\newboolean{biblio}
\newboolean{biblio2}
\newboolean{si}
\newboolean{exp}

\setboolean{showfigures}{true} 
\setboolean{biblio}{true} 
\setboolean{biblio2}{false} 
\setboolean{si}{true} 
\setboolean{exp}{true} 

\journal{}
\biboptions{numbers,sort&compress}

\begin{document}

\def\sectionautorefname{Section}
\def\subsectionautorefname{Subsection}
\def\subsubsectionautorefname{Subsection}

\begin{frontmatter}

\title{Identifying Dealbreakers and Robust Policies for the Energy Transition Amid Unexpected Events}

\author[1]{Diederik Coppitters\corref{mycorrespondingauthor}}
\ead{diederik.coppitters@uclouvain.be}

\author[2]{Gabriel Wiest}

\author[2,3]{Leonard G\"oke}

\author[1]{Francesco Contino}

\author[2]{Andr\'e Bardow}

\author[2]{Stefano Moret}

\cortext[mycorrespondingauthor]{Corresponding author}

\address[1]{Institute of Mechanics, Materials and Civil Engineering (iMMC), Universit\'e catholique de Louvain (UCLouvain), Place du Levant, 2, 1348 Louvain-la-Neuve, Belgium}

\address[2]{Energy \& Process Systems Engineering, Department of Mechanical and Process Engineering, ETH Z\"urich, Switzerland}

\address[3]{Reliability and Risk Engineering, Institute of Energy and Process Engineering, ETH Z\"urich, Switzerland}

\begin{abstract}
Disruptions in energy imports, backlash in social acceptance, and novel technologies failing to develop are unexpected events that are often overlooked in energy planning, despite their ability to jeopardize the energy transition. We propose a method to explore unexpected events and assess their impact on the transition pathway of a large-scale whole-energy system. First, we evaluate unexpected events assuming ``perfect foresight", where decision-makers can anticipate such events in advance. This allows us to identify dealbreakers, i.e., conditions that make the transition infeasible. Then, we assess the events under ``limited foresight" to evaluate the robustness of early-stage decisions against unforeseen unexpected events and the costs associated with managing them. A case study for Belgium demonstrates that a lack of electrofuel imports in 2050 is the main dealbreaker, while accelerating the deployment of renewables is the most robust policy. Our transferable method can help policymakers identify key dealbreakers and devise robust energy transition policies.
\end{abstract}


\end{frontmatter}

\clearpage
\section{Introduction}

Energy planning models involve numerous inputs that are inherently uncertain, such as technology and resource availability, energy commodity prices, and demand levels~\cite{moret2017characterization}. Modellers represent this uncertainty---if they consider it at all---with probabilities or uncertainty ranges~\cite{Mavromatidis2018} and generate results based on them, following a predict-then-act approach~\cite{yue2018review}. However, this approach tends to underestimate the full spectrum of possible futures~\cite{pye2015integrated}, leading to overconfidence in the selected course of action~\cite{lempert2013making}. This underestimation occurs because uncertainty characterizations focus on the ordinary and predictable, and overlook, for instance, unexpected events~\cite{coppitters2023optimizing}. Here, we define unexpected events as events that are largely mispredicted or unpredicted, significantly deviate from expectations, and persist long enough to influence strategic decisions, such as social resistance blocking wind power expansion~\cite{enevoldsen2016examining}, a disruption in resource supplies due to geopolitical tensions~\cite{kitamura2017energy}, or the failed breakthrough of anticipated technologies like nuclear Small Modular Reactors (SMRs)~\cite{vanatta2023technoeconomic}. Details about the terminology are provided in Supplementary~Note~1. Although rare, unexpected events can substantially disrupt the course of action and transform dependencies into vulnerabilities~\cite{iea2024}. For instance, if an anticipated resource like green hydrogen falls short, initial investments in hydrogen infrastructure could become obsolete~\cite{neumann2023potential}. These unexpected events should, therefore, not be ignored in the decision-making process~\cite{mccollum2020energy}. Yet, their rarity and the limited data on their likelihood make them unsuitable for conventional risk-informed approaches, as the probability of unexpected events is deemed so negligible that standard uncertainty propagation techniques fail to recognize them as viable scenarios~\cite{ioannou2017risk}. Therefore, methods are needed to assess the role of unexpected events in energy system planning~\cite{veeramany2016framework}.

Existing studies primarily use Story-and-Simulation (SAS) approaches to assess the vulnerability of energy systems to a specific unexpected event~\cite{grubler2018low, anable2012modelling}. These assessments lead to proposed mitigation strategies in a specific context, such as adopting integrated energy markets to reduce risks related to supply disruptions~\cite{di2024natural} or deploying existing low-carbon technologies to avoid reliance on unicorn technologies that may never materialize~\cite{heuberger2018impact}. Although valuable, SAS approaches rely on predefined scenarios~\cite{fortes2015long}, which limits their capacity to guide decision-making across a broad spectrum of unexpected-event scenarios~\cite{ku2024grand}. In the context of decision-making across a broad spectrum of scenarios with unknown likelihoods, methodologies from Decision Making under Deep Uncertainty (DMDU) are gaining traction~\cite{marchau2019decision}. Unlike traditional predict-then-act approaches, where energy system planners typically agree on assumptions about the input parameter uncertainties, DMDU focuses on agreeing on decisions by exploring a variety of possible futures, identifying vulnerabilities, and proposing robust alternatives~\cite{stanton2021decision}. Despite its potential, the application of DMDU methods in energy system planning is scarce, applied only for uncertainties within predictable ranges~\cite{paredes2024characterizing, groves2020benefits, haas2023dealing}, and seemingly unexplored in combination with unexpected events.

An additional drawback of existing studies assessing unexpected events in energy system planning is that nearly all the employed energy system models operate under a perfect foresight approach~\cite{mccollum2020energy}. This unrealistic approach assumes complete knowledge about the future when optimizing investment decisions, allowing early-stage decisions in the energy transition to be tailored for a specific unexpected event predicted far in advance~\cite{limpens2024energyscope}. Myopic pathway optimization models are better suited for this context, as they enable sequential decision-making with limited foresight~\cite{babrowski2014reducing}. In this more realistic setting, early-stage investment decisions are made without knowledge of an unexpected event that may later arise. Yet, myopic decision-making models are relatively rare~\cite{poncelet2016myopic, nerini2017myopic}, with only a handful applications incorporating SAS-based unexpected event assessments~\cite{heuberger2018impact, li2023long}.

In this paper, we propose a method to assess unexpected events in energy system planning (\autoref{fig:schematic_methods}). The method enables an open exploration of unexpected-event scenarios and leverages both perfect and myopic foresight to identify the dealbreakers for the energy transition and emphasize the significance of early-stage decisions. Specifically, we examine combinations of the main unexpected events that could jeopardize the transition, including unrealized breakthroughs in unicorn technologies, sharp deteriorations in exchange rates with other economies driving up the cost of imported technologies, geopolitical tensions disrupting energy imports, and various forms of social resistance: resistance to building renovations, to mobility changes, to the widespread deployment of renewable technologies and to nuclear power. The impact of these unexpected-event scenarios on achieving an energy transition within climate targets at a reasonable cost is assessed using an energy transition pathway optimization model within a whole-energy context---considering power, heating, mobility, and non-energy demand~\cite{contino2020whole}. The model is released as open-source and can be accessed on GitHub (\url{https://github.com/DCoppitters/EnergyScope_pathway_unexpected_events/releases/tag/v1.0.0}), while all our results are freely available on Zenodo (\url{https://doi.org/10.5281/zenodo.14845623}).

The evaluation of these unexpected-event scenarios follows a two-stage approach: First, we evaluate the unexpected-event scenarios under perfect foresight, i.e., an ideal case where the decision-maker is aware of the unexpected-event scenario from the start of the transition. From scenarios that result in a failed energy transition, i.e., the infeasible scenarios, we identify dealbreakers: conditions that make the transition infeasible, even when decision-makers are aware of unexpected events in the distant future and can make early-stage decisions accordingly. The feasible scenarios, however, are not guaranteed to remain feasible in a more realistic setting---a myopic foresight---as decision-makers may not make the right mitigating decisions early enough. Therefore, in the second stage, we reassess the feasible scenarios using the more realistic myopic foresight approach. Here, scenarios are revealed in 5-year intervals, and early-stage decisions are made without knowledge of events that will occur later in the transition. We impose various early-stage policies to evaluate how these decisions affect the robustness of the energy transition to unforeseen unexpected events arriving later on, and the costs associated with managing them.

\ifthenelse{\boolean{showfigures}}{
\begin{figure}[h!]
\centering
\includegraphics[width=0.68\textwidth]{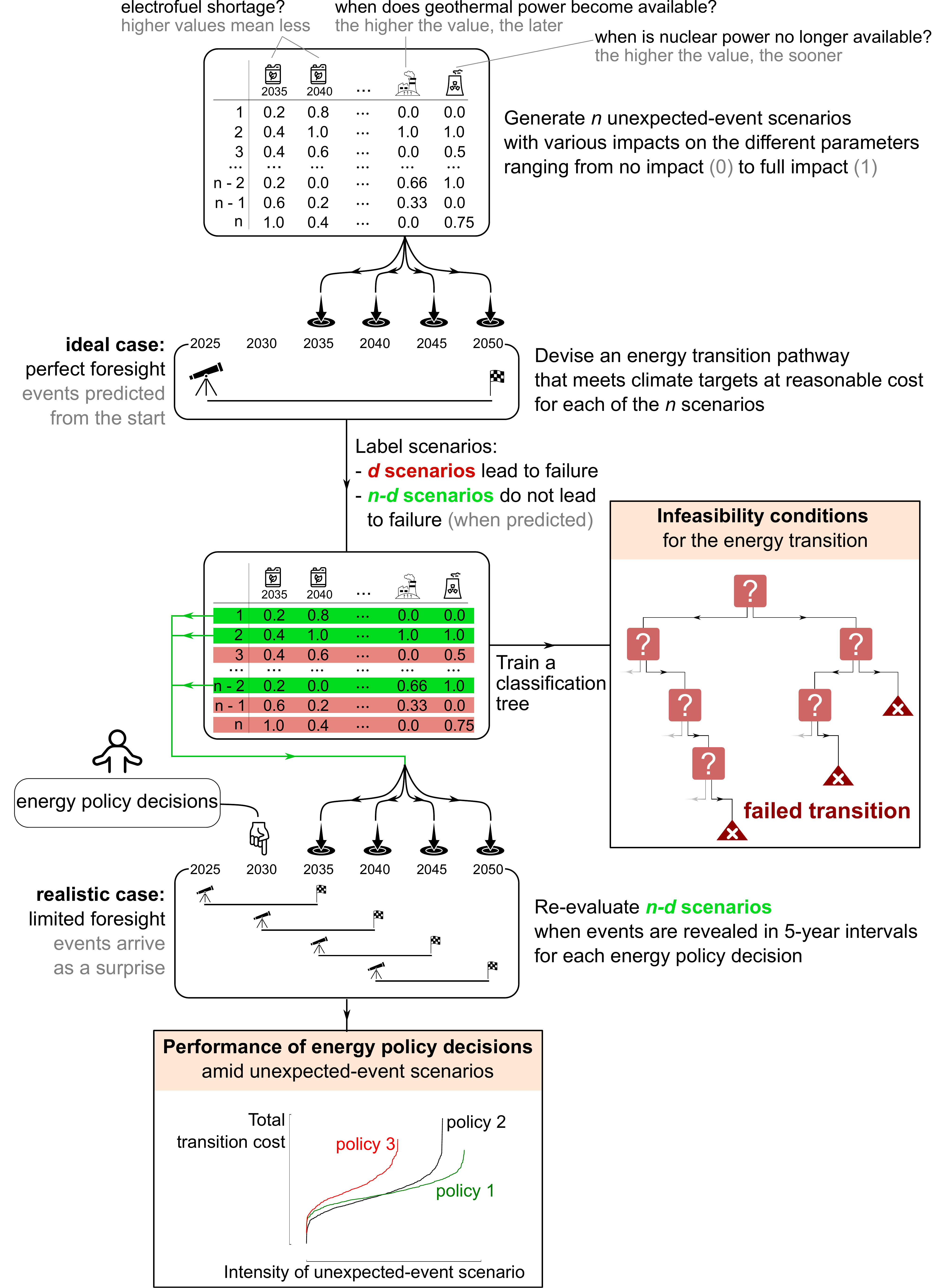}
\caption{Scenarios are generated by exploring the unexpected-events space and evaluated using an energy transition pathway optimization model with perfect foresight. This ideal setting optimizes the energy transition for each scenario starting with full knowledge on the future. In other words, if the pathway optimization model fails to find a solution, the scenario will inevitably lead to a failed transition. By labeling the scenarios based on success or failure, a classification tree can derive the infeasibility conditions. The $n-d$ feasible scenarios are then reassessed with limited foresight---a more realistic setting in which scenarios are revealed in 5-year intervals, and decisions are made without knowing future events. Thus, scenarios that are feasible with perfect foresight might become infeasible with limited foresight due to an early-stage decision that could hinder the ability to address unforeseen unexpected events that arise later. By starting from specific energy policy decisions (e.g., accelerated deployment of renewables, early phase-out of nuclear power), these scenarios are evaluated to determine how many unexpected-event scenarios are managed and the associated costs. This approach allows policymakers to assess the performance of energy policy decisions in relation to unexpected events.}
\label{fig:schematic_methods}
\end{figure}
}{}

In summary, our approach is not limited by specific narratives on unexpected events, but explores a wide range of possibilities. Moreover, the method leverages both the perfect foresight and myopic approaches to identify the dealbreakers for the energy transition, and to evaluate the performance of early-stage energy policy decisions in the context of unexpected events, respectively. The approach can provide policymakers with insights into critical vulnerabilities and helps identify robust energy policies against unexpected events. 

While this method can be applied to any region, we used Belgium as a case study due to its interesting challenges related to a limited renewable energy potential and a high population density~\cite{limpens2024energyscope}.

\section*{Results}

\subsection{Dealbreakers for the energy transition}
\label{sec:res:vulnerabilities_and_necessary_conditions}

The parameters representing unexpected events are defined based on unexpected events identified in the literature (detailed in Supplementary~Note~2). Their values range from 0 to 1, indicating the intensity of each event. The parameters are organized in 10 groups, i.e., related to the availability of imported electrofuels, biofuels, and electricity; societal resistance to building renovations, mobility changes and the deployment of PhotoVoltaics (PV), onshore wind and offshore wind; increased energy demand and increased costs of imported technologies. The impact on these 10 groups are independently sampled for the years 2035, 2040, 2045, and 2050, yielding a total of 40 parameters. An additional 6 parameters capture the availability of nuclear Small Modular Reactors (SMRs), Direct Air Capture (DAC), Carbon Capture and Storage (CCS), geothermal power, ammonia-fired Combined Cycle Gas Turbines (CCGTs), and ammonia cracking. Here, the values determine if and when these technologies become viable. The last parameter addresses whether and when a nuclear phase-out occurs.

The unexpected-event scenarios---combinations of impacts on the 47 parameters---are initially evaluated in the energy transition pathway optimization model using a perfect foresight approach. Following the evaluation, the scenarios are labeled as either successful or failed, depending on whether the climate targets can be achieved within a reasonable cost increase. Specifically, a failed transition indicates that the limit on the GreenHouse Gas (GHG) emissions is breached or the total transition cost to meet the GHG emission limit is excessive (details on how an excessive cost is defined are provided in the Experimental Procedures). The perfect foresight method ensures that if the pathway optimization model fails to identify a solution for a given scenario, no solution exists within the decision space, even when the unexpected-event scenario is known from the beginning of the transition. Finally, we train a classification tree to identify the key scenarios leading to failure and quantify the feature importances to determine the most important unexpected events driving the transition to failure. 

The classification tree highlights the key scenarios leading to a failed energy transition (\autoref{fig:dt_scen_disc_specific}): If the electrofuel supply is cut off entirely in 2050, the energy transition will not be feasible. In scenarios where electrofuel imports are low in 2050, the transition fails if final energy demand rises. Even when the electrofuel supply is high in 2050, the transition can fail under a set of conditions in 2045: the complete loss of electrofuel imports, a high rise in final energy demand, and nuclear energy being no longer available. In summary, the following infeasibility conditions, i.e., the dealbreakers for the energy transition, are derived:
\begin{itemize}

\item No electrofuel imports in 2050;

\item A substantial increase in final energy demand by 2050, coupled with low electrofuel imports;

\item No electrofuel imports by 2045, combined with the phase-out of nuclear energy and a sharp rise in final energy demand.

\end{itemize}

\ifthenelse{\boolean{showfigures}}{
\begin{figure}[h!]
\centering
\includegraphics[width=1\textwidth]{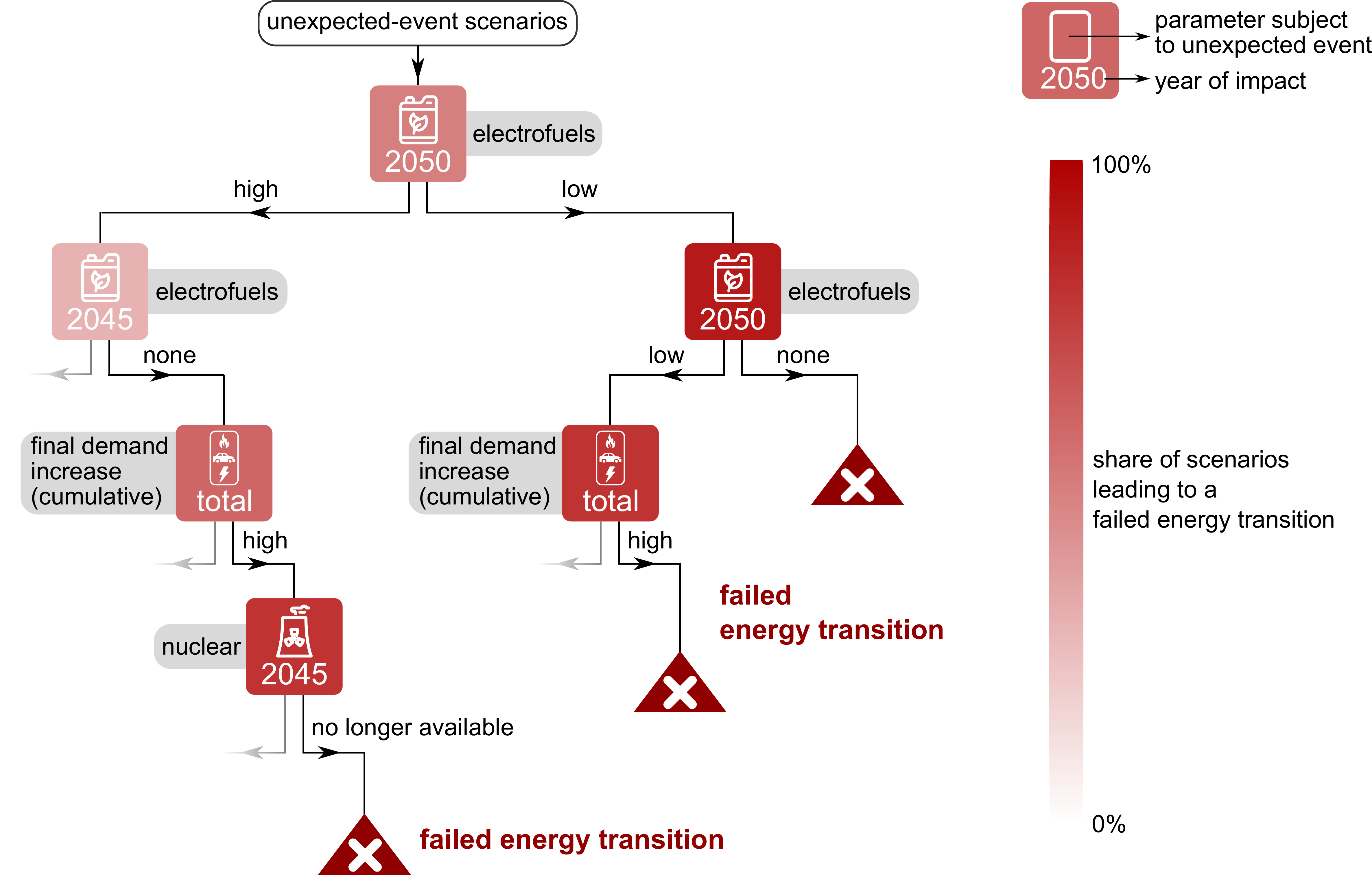}
\caption{Classification tree on the unexpected-event scenarios---labeled as either leading to a successful or failed energy transition under perfect foresight---to discover the infeasibility conditions for the energy transition. The classification tree, post-pruned to remove branches not leading to a failed energy transition, shows that the import of electrofuels in 2050 is a key enabler: if no electrofuels are imported in 2050, the energy transition will fail. In scenarios where the electrofuel imports are low in 2050, the transition fails if the final energy demand rises. Even when the electrofuel supply is high in 2050, the transition cannot be realized if none is available in 2045, combined with a phase out of nuclear power and a high rise of final energy demand by then. The shade indicates the share of failed energy transition scenarios at each node, with each split leading to a higher proportion in the right node.}
\label{fig:dt_scen_disc_specific}
\end{figure}
}{}

To make results more interpretable, we present a heatmap that illustrates the proportion of scenarios leading to failure with respect to two key parameters influencing the prediction accuracy of the classification tree: electrofuel availability in 2050 and 2045 (\autoref{fig:feature_matrix_efuels}). The heatmap highlights the dominant influence of electrofuel import availability in 2050 on the likelihood of a failed energy transition. When no imports are available in 2050, failure is certain. However, as availability increases, the failure likelihood decreases sharply, becoming nearly negligible once more than 40\% of the expected value (110~TWh/year out of 275~TWh/year) is accessible. This trend is largely unaffected by the availability of electrofuels in 2045, except in cases where no imports are available in that year. In such scenarios, the risk of failure remains high regardless of the availability in 2050.

\ifthenelse{\boolean{showfigures}}{
\begin{figure}[h!]
\centering
\includegraphics[width=0.8\textwidth]{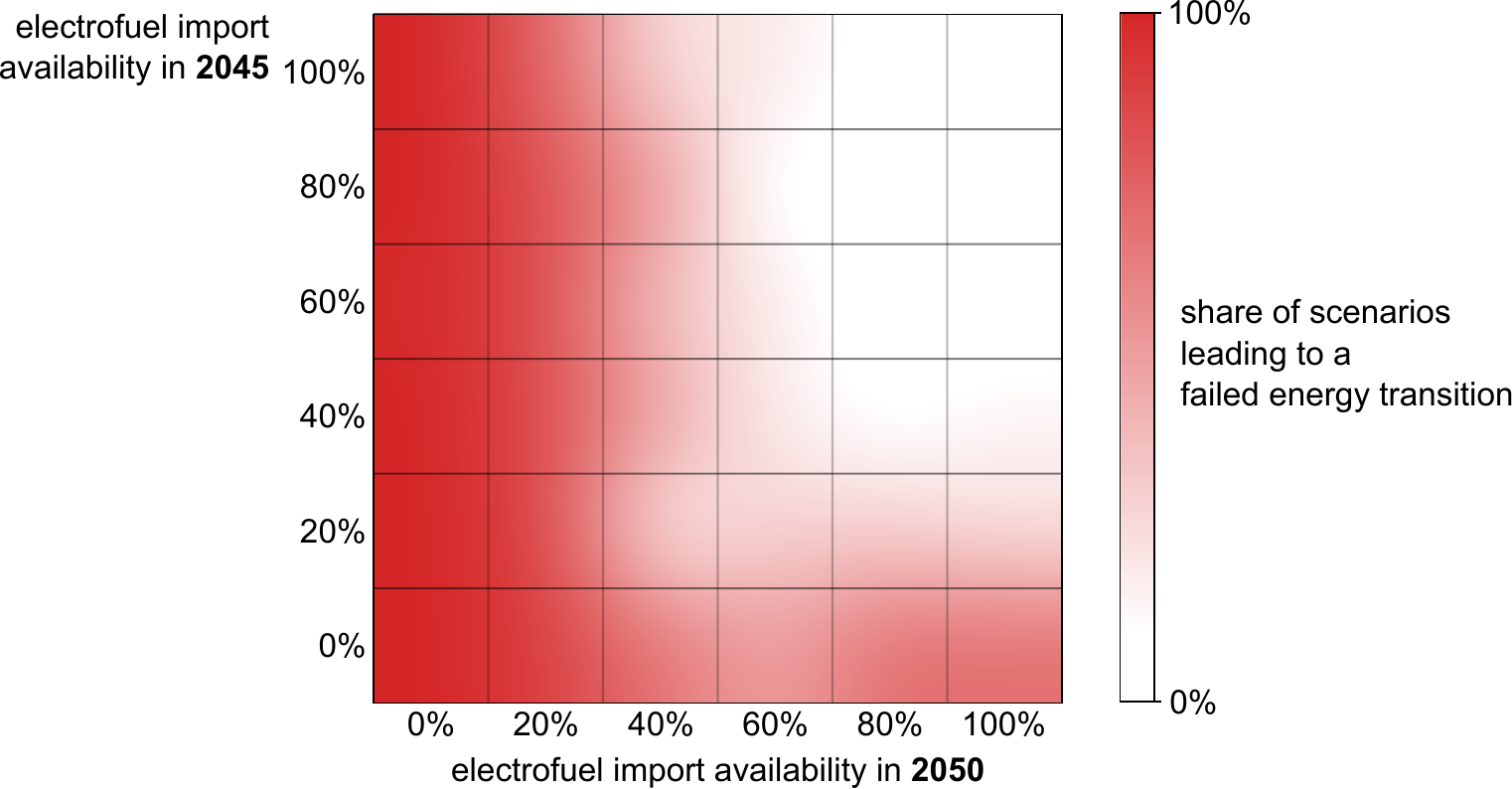}
\caption{
Heatmap illustrating the impact of electrofuel import availability in 2045 and 2050 on the likelihood of a failed energy transition. The risk of failure is almost certain with no electrofuels available by 2050. As import availability increases, the failure risk declines sharply, becoming minimal when availability exceeds 40\%. This trend remains consistent across all 2045 import levels, except when imports are zero, in which case the failure risk remains high at all 2050 availability levels. The cells are interpolated using a quadratic method to create a smooth transition between values.}
\label{fig:feature_matrix_efuels}
\end{figure}
}{}

\subsection{Robust energy policy decisions amid unexpected events}
\label{sec:res:vulnerability_earlystage_decisions}

The second stage of the method shifts from perfect foresight to myopic foresight to assess unexpected-event scenarios in a more realistic setting. We introduce early-stage decisions into the pathway model and evaluate, from a myopic perspective, how these initial choices affect the ability to handle unexpected events that emerge later in the transition. This approach better reflects reality, as energy policy decisions are often made early in the transition process with limited foresight, and pathways evolve as the future gradually unfolds. Only scenarios that are feasible under perfect foresight are considered, as those infeasible under perfect foresight would inherently be unmanageable under myopic foresight.

We incorporate various early-stage decisions into the myopic pathway optimization model for 2030, drawing from the REPowerEU chapter of Belgium's Recovery and Resilience Plan~\cite{beresplan}, the Federal Hydrogen Strategy~\cite{fedh2strat}, and the Belgian nuclear phase-out law~\cite{nuclearphaseoutlaw}. These decisions include: (1) accelerating the expansion of renewable energy technologies to their maximum potential; (2) phasing out nuclear power; and (3) making upfront investments in hydrogen-powered technologies. Additionally, we include a decision related to (4) delaying the deployment of renewable energy technologies, i.e., no changes to the energy system are made between 2025 and 2030 in this sector. While this may not be an energy policy that is currently on the table, recent years have been affected by delays due to political short-termism, public misunderstanding, and competing economic interests~\cite{stern2015we}, which makes it relevant to explore the potential consequences of continued inaction in the context of unexpected events. Detailed descriptions about the early-stage decisions are available in Supplementary~Note~6. Note that these energy policy decisions were not optimized, but rather are intended to show the effects of different policy directions and how they might impact the management of unexpected events. 

For each early-stage decision, investment decisions up to 2030 are fixed, while decisions from 2035 to 2050 are optimized depending on the unfolding unexpected-event scenario.  For the scenarios achieving a successful energy transition, the transition costs are arranged in ascending order, forming a cumulative curve of total transition cost relative to the number of unexpected-event scenarios that can be successfully managed when starting from a specific early-stage decision (\autoref{fig:cumcost}). Additionally, two reference cumulative cost curves are constructed for comparison: The myopic foresight baseline, where no investment decisions are exogenously introduced in 2030, and the perfect foresight baseline, which represents the transition costs attained by handling the unexpected-event scenarios under perfect foresight.

\ifthenelse{\boolean{showfigures}}{
\begin{figure}[h!]
\centering
\includegraphics[width=0.95\textwidth]{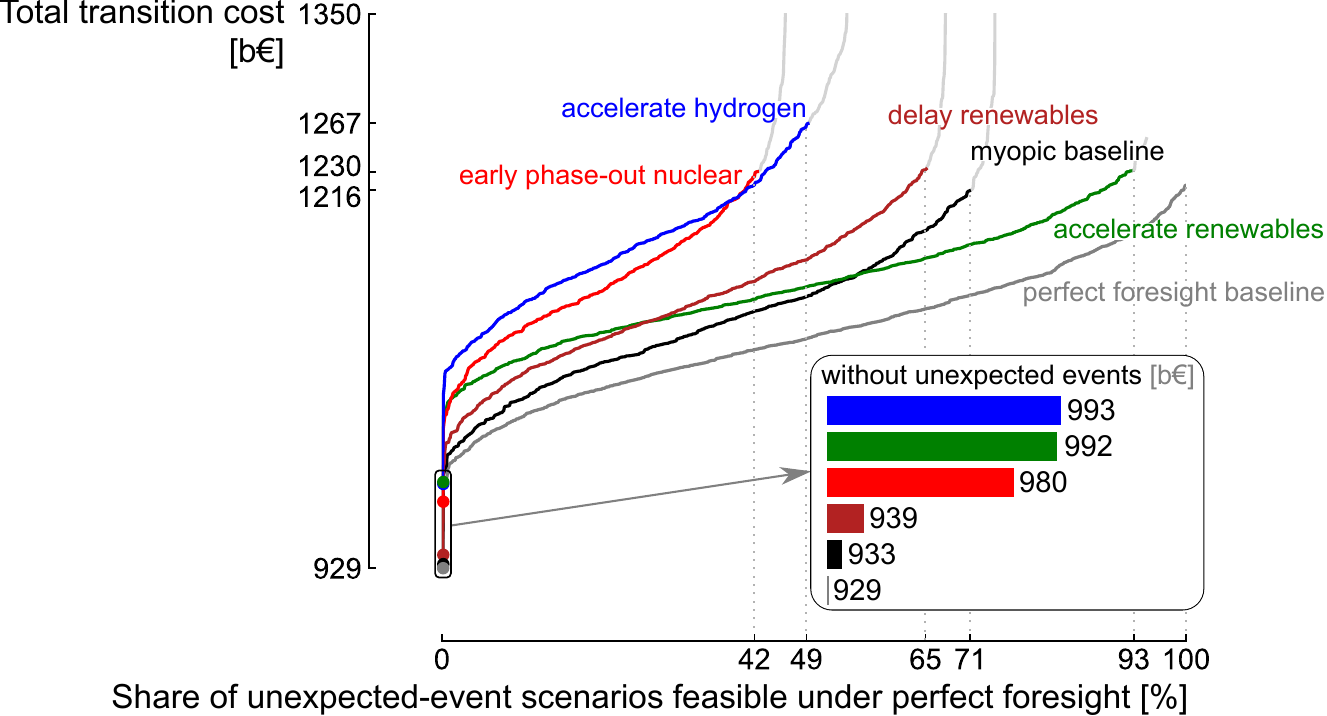}
\caption{The cumulative transition cost curves for different energy policy decisions taken in 2030 show that decisions accelerating the transition offer better robustness against unexpected events compared to the myopic baseline. Delaying investments in renewables increases vulnerability to unexpected events and makes them more costly to manage. Specifically, the policy decision to phase out nuclear power by 2030 significantly reduces robustness against unexpected events, handling only 42\% of the evaluated scenarios. The hydrogen route, which involves substantial deployment of hydrogen production and conversion technologies by 2030, is the most costly among all evaluated decisions and is less robust against unexpected events than the baseline decision due to greater reliance on electrofuel production and imports later in the transition. The perfect foresight approach (grey curve) represents the theoretical lower bound for these curves, as perfect foresight allows managing unexpected events at the lowest possible cost. The colored lines represent the cumulative cost curves, where the costs for managing unexpected-event scenarios remain within a reasonable range. The grey continuation of these curves illustrates how the costs further evolve at unreasonable rates.}
\label{fig:cumcost}
\end{figure}
}{}

The results show that the energy policy favoring an accelerated deployment of renewable energy technologies is better equipped to manage a wider range of unexpected-event scenarios compared to the myopic baseline decision (\autoref{fig:cumcost}), which deploys only moderate renewable energy capacities by 2030. However, for low to moderate unexpected-event scenarios, the total transition costs are lower when starting from the myopic baseline decision than from the accelerated renewables policy decision. In these cases, the baseline decision curve aligns more closely with the perfect foresight curve, while the accelerated renewables curve starts higher due to substantial upfront investments in renewable technologies. The accelerated renewables policy oversizes renewable capacity early on, missing opportunities to delay these investments and benefit from lower future investment costs. Consequently, it incurs higher transition costs in the absence of unexpected events (992~b\EUR compared to 933~b\EUR).

Delaying the further adoption of renewable energy technologies until after 2030 slightly increases the total transition cost (939~b\EUR) when no unexpected events are present during the transition, and it increases the vulnerability to unexpected-event scenarios compared to the myopic baseline. The two more specific energy policies---phasing out nuclear power by 2030 and accelerating the deployment of hydrogen-powered technologies---perform worse than the myopic baseline. Phasing out nuclear power substantially weakens the system's ability to handle unexpected events, managing only 42\% of the scenarios that were feasible under perfect foresight, due to Belgium's already limited options for low-carbon power supply and the increased reliance on imports as a consequence. Similarly, the hydrogen route results in higher transition costs and lower robustness to unexpected events compared to the baseline decision.

As expected, the myopic pathways are unable to achieve a successful energy transition across all scenarios that are feasible under perfect foresight: only 71\% of these scenarios remain feasible when using the myopic baseline approach. To understand why the myopic baseline approach fails in scenarios where perfect foresight succeeds, we categorize the feasible scenarios under perfect foresight into those that remain feasible under myopic foresight (71\%) and those that failed (29\%). Using this labeled dataset, we generate a classification tree to identify the three most important features influencing success or failure by quantifying their importance scores (\autoref{tab:feature_importances_}). These scores reflect the relative contribution of each parameter to the prediction accuracy of the tree. The analysis reveals that cumulative social resistance to PV deployment throughout the transition---specifically, the sum of the resistances experienced in 2035, 2040, 2045, and 2050---plays a critical role, with an importance score of 0.64. This parameter is followed by the loss of electrofuels in 2045 (importance score of 0.19) and the rise in total final energy demand (0.17). In the myopic baseline approach, PV deployment progresses at a moderate rate. However, due to the potential unforeseen arrival of social resistance to PV deployment---a factor that would have been anticipated and mitigated with perfect foresight---PV capacity may ultimately fall short. This shortfall, coupled with potential losses in electrofuel imports and rise in energy demands, limit the renewable energy levels required to meet final GHG emission targets by the end of the transition.

The analysis is extended to the myopic pathways starting from the different early-stage decisions (\autoref{tab:feature_importances_}). For the pathway starting with a delay in renewable power deployment, the success of the energy transition is most vulnerable to resistance against PV deployment and nuclear power, which jeopardizes reaching the necessary low-carbon power supply later on. A similar pattern is observed when starting with an accelerated rollout of hydrogen-powered technologies. In this case, the demand of hydrogen in cogeneration makes PV deployment critical in the early stages to produce the hydrogen, as electrofuel imports are still limited. When nuclear power is phased out early, PV deployment is accelerated to replace it, reducing vulnerability to resistance to PV deployment later. By the end of the transition, PV is fully deployed, and electrofuel imports increase to compensate for the loss of nuclear power. Therefore, this pathway is more vulnerable to increasing energy demand (0.27) and losses in electrofuel imports (0.39 for 2050, 0.34 for 2045). Finally, maximizing renewable power deployment from the start mitigates the impact of PV resistance, as deployment reaches full capacity early. The main vulnerability in this pathway lies, similarly to the early phase-out of nuclear power, in rise in final energy demand and electrofuel import losses.

\begin{table}[!ht]
    \centering
\resizebox{\textwidth}{!}{%
        \begin{tabular}{>{\raggedleft\arraybackslash}p{0.25\textwidth} | >{\centering\arraybackslash}p{0.13\textwidth} | >{\centering\arraybackslash}p{0.13\textwidth} >{\centering\arraybackslash}p{0.13\textwidth} >{\centering\arraybackslash}p{0.13\textwidth} >{\centering\arraybackslash}p{0.13\textwidth} >{\centering\arraybackslash}p{0.13\textwidth} >{\centering\arraybackslash}p{0.13\textwidth} >{\centering\arraybackslash}p{0.13\textwidth}}
    \hline \hline
         & &  \multicolumn{7}{c}{\textbf{feature importance scores} (relative contribution to success/failure prediction accuracy)} \\ \cline{3-9}
       \textbf{myopic pathway} &  \textbf{share of successful scenarios} &  resistance against PV deployment: 2035-2050 & electrofuels import loss 2040  & electrofuels import loss 2045  & electrofuel import loss 2050  & rise in total final energy demand  & resistance against nuclear power  &  resistance against PV deployment in 2035 \\ \hline 
         early phase-out nuclear  & 42\% & ~ & ~ & 0.34 & 0.39 & 0.27 & ~ & ~ \\[1ex]
         accelerated hydrogen  & 49\% & ~ & 0.33 & ~ & ~ & 0.24 & ~ & 0.43 \\[1ex]
         delay renewables  & 65\% & 0.57 & ~ & ~ & ~ & ~ & 0.13 & 0.30 \\[1ex]
         myopic baseline  & 71\% & 0.64 & ~ & 0.19 & ~ & 0.17 & ~ & ~ \\[1ex]
         accelerate renewables  & 93\% & ~ & ~ & 0.28 & 0.26 & 0.46 & ~ & ~ \\
         \hline \hline
    \end{tabular} }
    \caption{To understand why myopic pathways fail in scenarios where perfect foresight succeeds, we train a classification tree for each myopic pathway. The tree classifies scenarios where perfect foresight succeeded, labeling them as 0 if the myopic pathway also succeeded and 1 if it failed. Feature importance scores are then quantified to assess the relative significance of each feature in determining success or failure. To balance interpretability and accuracy, the tree is limited to three unique features. Results reveal that pathways lacking early acceleration in PV deployment are more susceptible to resistance against PV adoption during the transition. In contrast, the accelerated renewables scenario, characterized by early maximization of PV and wind deployment, faces vulnerabilities related to increases in final energy demand and electrofuel import losses.}
    \label{tab:feature_importances_}
\end{table}

\section{Discussion}

Our method introduces two novelties in assessing unexpected events in energy system planning: (1) the exploration of unexpected-event scenarios in a pathway context and (2) the use of both perfect foresight and myopic foresight to identify dealbreakers and emphasize the significance of early-stage decisions in the energy transition. 

First, unlike traditional approaches that rely on predefined scenarios typically focused on ordinary and predictable events~\cite{mccollum2020energy}, our method enables an exploration of possibilities. The exploration method remains tractable due to the computational efficiency of both the whole-energy system pathway optimization model and the exploration method. Specifically, myopic pathway evaluations take approximately 6~min on a 2.4~GHz 4-core machine and the exploration method requires only a few hundred scenario evaluations for sufficiently accurate cumulative cost curves, as detailed in Supplementary~Note~4. Moreover, the method supports parallel processing across multiple Central Processing Units (CPUs), which further reduces computation time.

Second, the method leverages both perfect foresight and myopic approaches to evaluate unexpected events within the whole-energy system pathway model. In the perfect foresight approach---the ideal case---decision-makers are fully aware of potential unexpected events from the start. This approach helps identify the infeasibility conditions which are visualized and interpreted using a classification tree. On the other hand, the more realistic myopic foresight approach introduces the element of surprise, with decision-makers gradually becoming aware of future events over time when starting from a specific energy policy decision. By combining these approaches, policymakers gain clear insights into critical vulnerabilities and robust early-stage policy decisions for managing unexpected events.

Applying the method to the case study highlighted its effectiveness: In the context of Belgium's energy transition, the classification tree identifies the loss of electrofuel imports in 2050 as a primary dealbreaker for success. Based on this finding, specific recommendations can be made to mitigate this risk. In this case, we recommend securing long-term trade agreements with key exporting countries and employing a contractual framework similar to early Liquified Natural Gas (LNG) markets---featuring bilateral agreements, long-term commitments and take-or-pay clauses~\cite{dejonghe2023natural}. Additionally, diversifying suppliers, routes, and carriers should be prioritized to mitigate supply security issues, despite potential higher costs. 

The evaluation of proposed energy policies with the myopic approach indicates that decisions prioritizing the acceleration of the transition are better equipped to handle a wider range of unexpected-event scenarios. Conversely, delaying the adoption of renewable energy increases vulnerability to unexpected events. While the benefits with respect to cost of accelerating the transition have been highlighted in previous studies~\cite{way2022empirically, de2024climate}, our results strengthen this direction by focusing on unexpected-event management. Moreover, these results likely apply to other regions where energy imports play a significant role (e.g., Germany~\cite{sun2024indispensable}), but further research is needed to verify whether similar dynamics and challenges exist elsewhere.

Applying this method to a larger energy transition model, such as a multi-region model connecting different areas (e.g., European Union member states)~\cite{goke2021graph}, could provide comprehensive insights into regional and systemic vulnerabilities to unexpected events. For example, our method could identify the dealbreakers for various European energy strategies. For instance, a continental-scale energy supply strategy would be the least-cost option but relies on unequally distributed generation across Europe---peripheral regions like Ireland and Albania would generate several times more than their demand to export to regions heavily dependent on imports, such as Belgium and Germany~\cite{trondle2020trade}. In contrast, a regional-scale supply, where regions are self-sufficient and isolated, would reduce dependency on imports but lead to significant cost disparities between regions. More broadly, in the context of energy security, our method could identify the risks associated with dependency on fossil gas imports and reveal how disruptions in one country might affect the entire EU's energy stability~\cite{lau2022europe}. These insights could contribute to developing a more unified and robust EU energy strategy, enhancing robustness against unexpected events through coordinated policies.

Nevertheless, our work has several limitations. Due to the extensive number of unexpected events considered and the different stages at which they could occur during the transition, exploring all combinatorial combinations (\SI{\approx e35}{}) is infeasible. Therefore, we rely on exploration approaches to generate a representative set of scenarios to represent the space, as is commonly done in DMDU~\cite{paredes2024characterizing}. Despite the sufficient accuracy of the exploration approach in the classification tree and cost curves (detailed in Supplementary~Note~4), it remains an approximation of the entire space. Moreover, only negative unexpected events are considered, while the energy transition could potentially benefit from positive unexpected events (e.g., abundant electrofuel supply). Combining the analysis of both negative and positive unexpected events in the analysis opens the door to evaluating the antifragile nature of energy policy decisions, where certain decisions not only remain robust against negative events but also take advantage of positive ones~\cite{coppitters2023optimizing}, providing a basis for future work.

Navigating the decision space to manage unexpected events while staying within climate targets is only possible by increasing the total transition costs. Although higher transition costs can be justified to achieve climate goals, there is a limit to this tolerance~\cite{steg2015understanding}. Eventually, a tipping point will be reached where excessive costs are no longer justifiable, forcing governments to accept higher GHG emissions to keep costs manageable. Determining when transition costs become excessive is subjective, and defining a precise threshold involves significant assumptions. In our study, this threshold is identified by the sharp, exponential increase in the cumulative cost curve, which provides cost tolerances in line with the 10\% to 30\% relaxation range commonly used in Modeling to Generate Alternatives (MGA) approaches~\cite{pickering2022diversity, lombardi2020policy}.

Additionally, our method assumes a linear decrease in GHG emissions allowed~\cite{limpens2024energyscope}. This approach limits the motivation for decision-makers in the early stages of the transition to push beyond the initial GHG emission limits, even though it might be beneficial to accelerate the transition early on. To address this limitation, a carbon budget could be assigned to the transition. While feasible under perfect foresight---where the pathway is optimized in one stage---a budget is incompatible with a myopic approach, which does not optimize the full transition at once and requires intermediate GHG emission targets~\cite{limpens2024energyscope}. To integrate a carbon budget within a myopic optimization framework, a combination of the pathway optimization model and reinforcement learning could be considered, where an agent is trained through interactions with its environment and repeats the entire transition with different sequences of actions and states to develop an optimized policy~\cite{rixhonphdthesis, perera2021applications}. This approach could also address another limitation of our method, where energy policy decisions must be predefined in the model. However, determining the agent's rewards could be challenging, as our method involves a range of unexpected-event scenarios and a cumulative cost curve, rather than minimizing costs for a single scenario. This approach, combined with positive unexpected events and antifragility metrics to reward the agent as mentioned earlier, will be considered in future work.

Even in its current state, our method facilitates an open exploration of possibilities, identifies infeasibility conditions, and assesses the performance of energy policy decisions in the context of unexpected events. The results provide policymakers with clear insights into critical vulnerabilities, helps formulate robust energy policies to address these events, and reveals any extra costs involved.

\ifthenelse{\boolean{exp}}{

\section{Methods}
\label{sec:methods}

\subsection{Resource availability}

\subsubsection*{Lead Contact}
\noindent
Further information and requests for resources and materials should be directed to and will be fulfilled by the Lead Contact, Diederik Coppitters (\href{mailto:diederik.coppitters@uclouvain.be}{diederik.coppitters@uclouvain.be}).

\subsubsection{Materials availability}

\noindent
The large-scale whole-energy pathway model and the unexpected-event scenario evaluation script have been deposited to GitHub: \url{https://github.com/DCoppitters/EnergyScope_pathway_unexpected_events/releases/tag/v1.0.0}. All model results have been deposited to Zenodo: \url{https://doi.org/10.5281/zenodo.14845623}. 

\subsubsection{Data and code availability}

\noindent
All code and data associated with this study are available on GitHub: \url{https://github.com/DCoppitters/EnergyScope_pathway_unexpected_events/releases/tag/v1.0.0} and Zenodo: \url{https://doi.org/10.5281/zenodo.14845623}.

\subsection{Space of unexpected events}
\label{sec:extremes}

The parameters representing unexpected events are selected based on those highlighted in the literature (detailed in Supplementary~Note~2) and are presented in \autoref{tab:disruptive_events}. The parameters are organized in 10 groups: the availability of imported electrofuels, biofuels, and electricity; societal resistance to deploying PV, onshore wind and offshore wind, to low-temperature heating renovations, and to mobility changes; increased demand and rising import costs for specific technologies. Each group is assessed independently for the years 2035, 2040, 2045, and 2050, resulting in four parameters per group and a total of 40 parameters. These parameters are treated as discrete variables, each assigned values of 0\%, 20\%, 40\%, 60\%, 80\%, or 100\% of the maximum potential impact. The values vary independently across phases, enabling diverse scenario representations. For example, in one scenario, electrofuel supply might face a 60\% shortfall in 2035, no shortfall in 2040, a complete disruption in 2045, and a 40\% shortfall in 2050, reflecting potential geopolitical instability during the energy transition.

Technology availability impacts are modeled using seven additional parameters. For each unicorn technology, a parameter indicates when the technology becomes available: 2040 (0\% impact), 2045 (33\%), 2050 (67\%), or not at all (100\%). Similarly, nuclear power availability may be phased out by 2035 (100\% impact), 2040 (75\%), 2045 (50\%), or 2050 (25\%), or remain available throughout (0\%). Once phased out, the availability cannot be reversed.

In total, 47 parameters are considered: 40 for availability, resistance, demand and cost issues across the four phases (2035–2050), and seven for technology-specific disruptions. These include nuclear SMRs, DAC, CCS, geothermal power, ammonia-fired CCGTs, ammonia cracking, and conventional nuclear power phase-out.

\begin{table}[h!]
\centering

\resizebox{\textwidth}{!}{%
\begin{tabular}{llll}

        \hline \hline
Category & Number of parameters & Parameter values & Years of impact \\ 
 & (47 in total) & & \\ \hline
Availability imported resources: & & \\
\hspace{10pt} electrofuels & 4 (one per year) & [0, 20, 40, 60, 80, 100] & 2035, 2040, 2045, 2050 \\
\hspace{10pt} biofuels & 4 (one per year) & [0, 20, 40, 60, 80, 100] & 2035, 2040, 2045, 2050 \\
\hspace{10pt}  electricity & 4 (one per year) & [0, 20, 40, 60, 80, 100]  & 2035, 2040, 2045, 2050 \\ \\
Resistance to change: & & \\
\hspace{10pt} private mobility & 4 (one per year) &  [0, 20, 40, 60, 80, 100]  &  2035, 2040, 2045, 2050 \\
\hspace{10pt} low-temperature heating & 4 (one per year) &  [0, 20, 40, 60, 80, 100]  &  2035, 2040, 2045, 2050 \\
\hspace{10pt} PV deployment & 4 (one per year) &  [0, 20, 40, 60, 80, 100]  &  2035, 2040, 2045, 2050 \\
\hspace{10pt} wind onshore deployment  & 4 (one per year) &  [0, 20, 40, 60, 80, 100]  &  2035, 2040, 2045, 2050 \\
\hspace{10pt} wind offshore deployment & 4 (one per year) &  [0, 20, 40, 60, 80, 100]  &  2035, 2040, 2045, 2050 \\ \\
Exchange rate deterioration & 4 (one per year) &  [0, 20, 40, 60, 80, 100]  & 2035, 2040, 2045, 2050 \\ \\
Energy demand increase & 4 (one per year) &  [0, 20, 40, 60, 80, 100]  & 2035, 2040, 2045, 2050 \\ \\

Unicorn technologies & & &\\
\hspace{10pt} Ammonia-fired CCGT & 1 (value sets arrival) & $[0, 33, 67, 100]$ & 2040, 2045, 2050 \\
\hspace{10pt} Ammonia cracking & 1 (value sets arrival) & $[0, 33, 67, 100]$ & 2040, 2045, 2050 \\
\hspace{10pt} Direct air capture & 1 (value sets arrival) & $[0, 33, 67, 100]$ & 2040, 2045, 2050 \\
\hspace{10pt} Nuclear SMR & 1 (value sets arrival) & $[0, 33, 67, 100]$ & 2040, 2045, 2050 \\
\hspace{10pt} CCS & 1 (value sets arrival) & $[0, 33, 67, 100]$ & 2040, 2045, 2050 \\
\hspace{10pt} Geothermal power & 1 (value sets arrival) & $[0, 33, 67, 100]$ & 2040, 2045, 2050 \\ \\

Nuclear phase-out & 1 (value sets timing) & $[0, 25, 50, 75, 100]$ & 2035, 2040, 2045, 2050\\

        \hline
        \hline
\end{tabular} }
    \caption{Parameters subject to unexpected events, their potential impacts on expected values, and the years when these impacts occur. For example, the expected availability of imported electrofuels may be affected in 2035, 2040, 2045, and 2050 by reductions of 0\%, 20\%, 40\%, 60\%, 80\%, or 100\%, with each year potentially experiencing a different level of impact. For unicorn technologies, a single parameter per technology indicates their availability, with possible scenarios being: available in 2040 (0\% impact), 2045 (33\% impact), 2050 (67\% impact), or unavailable (100\% impact). Similarly, nuclear power may be phased out in 2035 (100\%), 2040 (75\%), 2045 (50\%), or 2050 (25\%), or it may remain available throughout the transition (0\%). Once phased out, nuclear power cannot be reinstated. }
    \label{tab:disruptive_events}
\end{table}

\subsection{Space exploration}
\label{sec:creationofdisruptiveeventsandtheirintroductioninpathwaymodel}

As the parameters subject to unexpected events can be impacted simultaneously during the transition, we generate scenarios combining impacts for each parameter. As the amount of combinations of potential scenarios is infeasibly large (\SI{\approx e35}{}), we perform a space exploration to generate a representative set of unexpected-event scenarios.

The method works as follows. Each parameter \( p_i \), where \( i = 1, 2, \ldots, 47 \), can take values in the range \([0, 1]\). To efficiently sample this high-dimensional space, we use the Sobol sequence, a low-discrepancy sequence that ensures more uniform coverage of the parameter space compared to purely random sampling methods~\cite{bratley1988algorithm}. The Sobol sequence \( \mathbf{S}_n = \{ \mathbf{s}_1, \mathbf{s}_2, \ldots, \mathbf{s}_n \} \) generates \( n \) quasi-random points in the unit hypercube \([0, 1]^{47}\). Each sample point \( \mathbf{s}_j \in \mathbf{S}_n \) is a 47-dimensional vector \( \mathbf{s}_j = (s_{j1}, s_{j2}, \ldots, s_{j47}) \), where \( s_{ji} \in [0, 1] \) for all \( i \).

The parameters related to resource availability, exchange rate, resistance to change, energy demand, and maximum potential installed capacity of renewable energy technologies can each assume one of six discrete values: \(\{0.0, 0.2, 0.4, 0.6, 0.8, 1.0\}\). For nuclear energy availability, the values include phase-out in 2035 (1), 2040 (0.75), 2045 (0.5), 2050 (0.25), or no phase-out (0). Similarly, the availability of unicorn technologies is represented by discrete values corresponding to availability in 2040 (0), 2045 (0.33), 2050 (0.67), or not at all (1). The continuous Sobol sample points are mapped to these nearest discrete values using the following formulas:

\begin{equation}
\tilde{s}_{ji} = \begin{cases} 
\min\left(\dfrac{\lfloor s_{ji} \times 6 \rfloor}{5}, 1\right) & \text{if } i \in T, \\
\min\left(\dfrac{\lfloor s_{ji} \times 5 \rfloor}{4}, 1\right) & \text{if } i \in N, \\
\min\left(\dfrac{\lfloor s_{ji} \times 4 \rfloor}{3}, 1\right) & \text{if } i \in U, \\
\end{cases}
\end{equation}

\noindent
where \( T \) represents the set of indices for parameters related to resource availability, exchange rate, resistance to change, energy demand, and maximum potential installed capacity of renewable energy technologies; \( N \) denotes the set of indices for the parameter related to nuclear availability; \( U \) is the set of indices for unicorn technology availability parameters; and \( \lfloor \cdot \rfloor \) is the floor function, which returns the greatest integer less than or equal to the input.

\subsection{Whole-energy system pathway model}
\label{sec:energyplanningmodel}

We use EnergyScope Pathway to optimize energy transition pathways for a whole-energy system, i.e., encompassing electricity, heating, mobility, and non-energy sectors~\cite{limpens2024energyscope}. The model minimizes the transition cost of the energy system while adhering to a linearly-decreasing GHG emission constraint. This is achieved by optimizing the deployment and usage of 111 technologies and 28 resources. CCS and DAC are limited to supplying CO$_2$ as a feedstock for producing carbon-based fuels (e.g., methane), rather than offsetting emissions from other technologies. 

EnergyScope Pathway operates with an hourly time resolution and employs a Typical Days approach to handle computational complexity. The model is structured around representative years selected every 5 years from 2020 to 2050, with each period between these years considered a phase. During each phase, investment and decommissioning decisions can be made to update the energy system layout. 

The model supports both perfect foresight and myopic evaluation approaches. In the perfect foresight approach, the model optimizes the entire transition process from start to finish with full knowledge of future conditions. In contrast, the myopic approach focuses on optimizing smaller time windows sequentially using a rolling horizon method. Each time window is optimized based on current information and assumptions, without considering future conditions beyond the immediate window.

We apply this model to the Belgian energy transition as a case study. The characteristics of the Belgian energy transition, including resources, technologies, and energy demands, are adopted from Limpens~et~al.~\cite{limpens2024energyscope}. While electrofuels and biofuels were initially assumed to be abundantly available, we restricted their availability to increase from \SI{20}{TWh \per year} in 2030 to \SI{275}{TWh \per year} in 2050, based on the most recent projections~\cite{belgium_import_hub}, with a linear increase between phases. In addition, the maximum potential installed capacities for renewable technologies are updated based on a more recent study~\cite{correa2023paths}. 

The model is released as open-source and is available on GitHub (\url{https://github.com/energyscope/EnergyScope_pathway}), along with an installation guide and documentation.

\subsection{Criteria for a successful energy transition}
\label{sec:definition_crit_event}

\noindent
In this work, the energy transition is considered to have failed when:

\begin{itemize}
\item The GHG emission limit is breached at any stage of the transition;
\item The total transition cost to meet the GHG emission limit is excessive.
\end{itemize}

\noindent
Although higher transition costs can be considered acceptable to achieve climate targets when unexpected events arrive, there is a limit to this tolerance~\cite{steg2015understanding}. There will be a tipping point where excessive costs are no longer justifiable, leading governments to accept higher GHG emissions to keep costs manageable. Determining when transition costs become excessive is subjective, and defining a precise threshold involves significant assumptions. Therefore, we identify this threshold a posteriori by evaluating the excess transition costs across all assessed unexpected events.

To assess when transition costs become excessive, we sort costs from low to high and normalize them against the optimal total transition cost achieved under perfect foresight, without any unexpected events. This creates a cumulative curve that tracks the total transition cost relative to the number of unexpected-event scenarios, arranged from the cheapest to the most challenging, resembling a rotated S-curve (\autoref{fig:cumcost}). For each point on the curve, we calculate the gradient. The cut-off point corresponds to where the gradient exceeds 1, meaning there is a cost increase of 1\%$_{\mathrm{abs}}$ for a 1\%$_{\mathrm{abs}}$ increase in the number of scenarios covered. Beyond this point, any further increase in costs is considered excessive. Note that the evaluation of the gradient begins when the percentage of unexpected-event scenarios covered reaches 10\%, starting sufficiently far along the plateau of the rotated S-curve, neglecting the rapid increase (with high gradients) at the beginning of the curve.

\subsection{Classification tree}
\label{sec:meth:scenariodiscovery}

We use a classification tree to identify the key features of the unexpected-event scenarios leading to a failed energy transition~\cite{baader2023streamlining}. After evaluating the unexpected-event scenarios in the pathway optimization model using perfect foresight, the scenarios are labeled with a binary target variable to indicate whether they resulted in a successful or failed energy transition. For this dataset, we apply a supervised machine learning technique using decision tree classification with the Classification and Regression Trees (CART) algorithm~\cite{breiman2017classification}. The tree is trained using the \textit{scikit-learn 1.3.0} package in Python~\cite{pedregosa2011scikit}. The dataset, including the 47 parameters subject to unexpected events, is supplemented with four additional parameters: cumulative resistance to PV deployment, onshore wind deployment, offshore wind deployment, and the cumulative rise in energy demand. These parameters correspond to the sum of their respective values for each impacted phase (2035, 2040, 2045, 2050).

To prevent overfitting and maintain interpretability, we limit the number of tree leaves by balancing interpretability and coverage scores ~\cite{gerst2013discovering}. The Interpretability score measures the number of unique features in the tree, while the coverage score reflects the proportion of data points correctly assigned. We also conduct $k$-fold cross-validation to assess the performance of the model on data not included in the training set~\cite{baader2023streamlining}. The classification tree scores are detailed in Supplementary~Note~4.

\subsection{Convergence assessment of cumulative cost curves}
\label{sec:meth:convergence}

The shape of cumulative transition cost curves for each energy policy decision depends on the number of unexpected-event scenarios evaluated. To determine whether the number of scenarios is sufficient, we analyze the convergence of two key metrics: the proportion of scenarios where a successful energy transition remains feasible, and the area under the curve. Detailed methods and results for these metrics are provided in Supplementary~Note~4.

\section*{Acknowledgments}

D.C. acknowledges the support of the Fonds de la Recherche Scientifique – FNRS [CR~40016260]. S.M. and G.W. acknowledge support from the Swiss National Science Foundation (SNSF) under Grant No. PZ00P2\_202117. A.B. acknowledges the support of the Swiss Federal Office of Energy through the SWEET consortium PATHFNDR under Grant No. SI/502259.

\section*{Author contributions}

Conceptualization, D.C., G.W., L.G., F.C., A.B., and S.M.; Methodology, D.C., G.W., L.G., and S.M.; Investigation, D.C.; Writing – Original Draft, D.C.; Writing – Review \& Editing, D.C., G.W., L.G., F.C., A.B., and S.M.; Funding Acquisition, D.C.; Resources, F.C. and S.M.; Supervision, S.M.

\section*{Declaration of interests}

A.B. served on review committees for research and development at ExxonMobil and TotalEnergies. A.B. and S.M. have ownership interests in firms that render services to industry, some of which may provide energy planning services. All other authors have no competing financial interests.

}


\clearpage

\appendixpageoff
\appendixtitleoff
\renewcommand{\appendixtocname}{Supplemental Information}
\begin{appendices}
\crefalias{section}{supp}

\setcounter{table}{0}
\renewcommand{\thetable}{S\arabic{table}}%
\setcounter{figure}{0}
\renewcommand{\thefigure}{S\arabic{figure}}%
\renewcommand{\thesection}{Supplementary Note \arabic{section}}%

\noindent
\textbf{\Large Supplemental Information \\ \\ Identifying Dealbreakers and Robust Policies for the Energy Transition Amid Unexpected Events} 

\noindent
\textbf{Diederik Coppitters, Gabriel Wiest, Leonard G\"oke, Francesco Contino, Andr\'e Bardow, Stefano Moret} \\

\noindent
The Supplemental Information begins with a discussion of the terminology used in the literature to describe unexpected events (\ref{sec:si:terminology}). It then provides further details on the unexpected events considered in this work (\ref{sec:si:unexpected_events}). Next, it presents the details of the classification tree (\ref{sec:si:tree_true}). The document then addresses the convergence and accuracy of the cumulative cost curves (\ref{sec:si:convergence_dt}). Following this, it examines the diversity in investment decisions across the unexpected-event scenarios, the correlation between these decisions, and their relationship with the unexpected events themselves (\ref{sec:si:designs_pf}). Finally, the document provides additional details on the early-stage energy policy decisions integrated into the energy transition pathway optimization model, evaluated against the unexpected-event scenarios (\ref{sec:si:energypolicydecisions}).

\clearpage

\section{Terminology used in the literature}
\label{sec:si:terminology}

The literature uses various terms to describe rare events with potentially disruptive impacts. In power system engineering, these events are commonly termed \textit{High-Impact Low-Probability events}~\cite{sperstad2020comprehensive}. McCollum~et~al.~\cite{mccollum2020energy} classify extremes relevant to energy modeling into three categories: \textit{transient events} (e.g., unexpected weather), \textit{disruptive drivers} (e.g., mass automation), and \textit{unexpected outcomes} (e.g., technological discoveries). The SWEET SURE scenario protocol~\cite{sweetsure} builds on this framework and defines sudden future events in a pathway as \textit{shocks}, including phenomena like population growth due to climate refugees, the reintroduction of nuclear power, and deteriorating exchange rates between Asia and the rest of the world, which raise import costs. They differentiate between transient shocks, which affect the pathway trajectory but do not change its end, and disruptive shocks, which do. Similarly, Heuberger~et~al.~\cite{heuberger2018impact} assess the impact of waiting for unicorn technologies to emerge on achieving climate targets and their costs. They refer to the arrival of such technologies as a \textit{disruptive event}. In contrast, Hanna and Gross~\cite{hanna2021energy} review how energy system models represent disruption and discontinuity, classifying \textit{disruptions} as temporary, not causing fundamental changes to equilibrium, while \textit{discontinuities} are irregular and lead to fundamental changes.

In our work, we focus on rare events that are largely mispredicted or unpredicted, significantly deviate from expectations, and persist long enough to influence strategic decisions---for example, over a 5-year phase of the energy transition. We refer to these events as \textit{unexpected events}, highlighting their rarity while deliberately avoiding assumptions about their potential impact. This distinction is important because terms like High-Impact Low-Probability events or disruptive events inherently imply significant consequences. In contrast, the unexpected events we examine may or may not have substantial impacts, which can only be determined after evaluating their influence within an energy pathway optimization model. By analyzing the model’s results, we assess whether these events lead to significant outcomes. Thus, to focus solely on their rarity before evaluating their impact, we intentionally use the term \textit{unexpected events}.

\clearpage

\section{Details on the unexpected events}
\label{sec:si:unexpected_events}

We look at unexpected events mentioned in the literature and identify how they could affect the parameters in the pathway model. Rather than assessing each unexpected event individually, we consider their combined impacts and generate scenarios encompassing a potential impact on each parameter. The events include geopolitical disruptions affecting the availability of imported resources~\cite{hanna2021energy}, the success or failure of novel technologies~\cite{heuberger2018impact}, resistance to behavioral and lifestyle changes such as building renovations, mobility changes, the adoption of renewable energy technologies or persistence of nuclear phase-outs~\cite{hanna2021energy}, exchange rate deteriorations increasing the costs of technology imports~\cite{sweetsure}, and rapid surges in energy demand~\cite{sweetsure}. Each event is detailed in the following paragraphs.

\paragraph{Availability of imported resources}

Given the reliance on imported energy, the energy transition pathways are exposed to risks of supply disruptions, particularly from geopolitical tensions~\cite{hanna2021energy}. We model scenarios of import losses for electrofuels, biofuels, and electricity, with disruptions occurring in five-year intervals. The impact ranges from no effect to a complete supply cut, across six discrete levels. We analyze the potential loss for each energy type separately, considering that different source regions might be available for the different energy imports: electricity is primarily sourced from neighboring countries, whereas electrofuels and biofuels will be mostly imported from overseas suppliers, such as Morocco and Oman for electrofuels and Brazil for biofuels~\cite{belgium_import_hub}. In total, twelve parameters are assessed, covering parameters per energy source (electrofuels, biofuels, and electricity) across each transition phase (2035, 2040, 2045, and 2050).

\paragraph{Resistance to change}

The pathway model includes society's resistance to change in various aspects of the transition. Specifically, a complete renovation of low-temperature heating technologies can only occur within a 15-year timeframe. This means that during each phase, a maximum of 33\% of the installed capacity can be replaced. Similarly, the entire passenger and freight mobility fleet can only be renewed over a period of 10 years, allowing for a maximum replacement of 50\% of the mobility fleet during each phase. Impacts associated with resistance to change vary during each phase of the transition, ranging from no impact (i.e., the expected inertia to change) to scenarios where no changes are permitted during that phase (indicating maximum impact). Additionally, the shift to renewable energy relies on widespread adoption of PV systems, onshore, and offshore wind turbines. However, their maximum potential installed capacity depends, among others, on socio-political acceptance. While social acceptance is a known challenge for wind energy, it also affects solar energy up to a similar level for solar farms with similar size to conventional wind turbines~\cite{cousse2021still}. We considered a reduced social acceptance for PV, onshore and offshore wind farms independently, fluctuating every five years from full acceptance to complete rejection, limiting further expansion in a given phase. Note that existing installations built in a previous phase are not forced to be decommissioned prematurely---it only limits future expansion.

\paragraph{Deterioration of exchange rate}
Belgium relies heavily on technology imports to decarbonize its energy system. Exchange rate fluctuations have historically increased the costs of these imports~\cite{sweetsure}. Asian economies are expected to dominate low-carbon technology production due to their lower prices. According to the Clean Energy Technology Observatory, the EU is vulnerable to a dependency during the energy transition on Asian markets for PV arrays~\cite{ceto_pv}, battery technologies and electric vehicles~\cite{ceto_battery}, heat pumps (mainly compressors and refrigerant supply)~\cite{ceto_hp} and hydrogen electrolyzers (mainly alkaline)~\cite{ceto_electrolysis}. To consider the potential negative impacts of expensive imports, we model a sudden deterioration in the exchange rate between Asian economies and Belgium. This impact is analyzed as a discrete variable, with cost increases on imported goods ranging from 0\% to 40\% during each phase of the transition, based on the low, middle, and high scenarios in the SWEET SURE scenario protocol~\cite{sweetsure}.

\paragraph{Demand increase}
Unanticipated increase in final energy demand can be driven by several factors, including population growth, strong economic growth and higher standards of living~\cite{sweetsure}. We consider the impact on the final energy demand as a discrete variable, potentially increasing the energy demand between 0\% and 15\% at each phase of the transition, potentially leading to a 60\% demand increase by 2050. 

\paragraph{Nuclear energy phase-out}
The future availability of nuclear energy remains uncertain due to socio-political concerns about safety, waste management, and the environmental impact of nuclear accidents, leading to public opposition and regulatory challenges~\cite{nguyen2018examination}. Therefore, nuclear power can be phased-out at any time during the transition. The impact of social resistance against nuclear energy is modeled by a single discrete parameter, which allows for scenarios of either a complete shutdown in 2035 (highest impact), 2040, 2045, or 2050, or the retention of full capacity (no impact).

\paragraph{Unicorn technologies}
Energy system planning models often rely on technologies that have proven effective on a demonstration scale but are not yet commercially deployed (i.e., unicorn technologies~\cite{heuberger2018impact}). This can result in proposed pathways for the energy transition that rely on waiting for these unicorn technologies to mature instead of investing in currently available options. This poses a significant risk because if these unicorn technologies fail to become available at a large scale, it could severely hinder the achievement of climate targets and increase costs significantly~\cite{heuberger2018impact}.
We assess the impact of unicorn technologies based on their availability either at the predicted time (no impact), later in the transition, or not at all (maximum impact). Based on the classification of unicorn technologies by Heuberger~et~al.~\cite{heuberger2018impact}, unicorn technologies fall between Technology Readiness Level (TRL)~3 and TRL~7 and we included ammonia-fired Combined Cycle Gas Turbines, ammonia cracking, Direct Air Capture, nuclear SMRs and advanced Carbon Capture and Storage based on the classification of the International Energy Agency~\cite{ieatrl}. While large-scale geothermal power plants are being developed commercially, their deployment in Belgium is limited, and few research projects have studied Belgium's geothermal potential~\cite{limpensthesis}. Therefore, we treat the availability of geothermal power (\SI{4}{GW_e}) similarly to unicorn technologies.

\clearpage

\section{Classification tree}
\label{sec:si:tree_true}

\begin{landscape}
\begin{figure}[h!]
\centering
\includegraphics[width=1\textwidth]{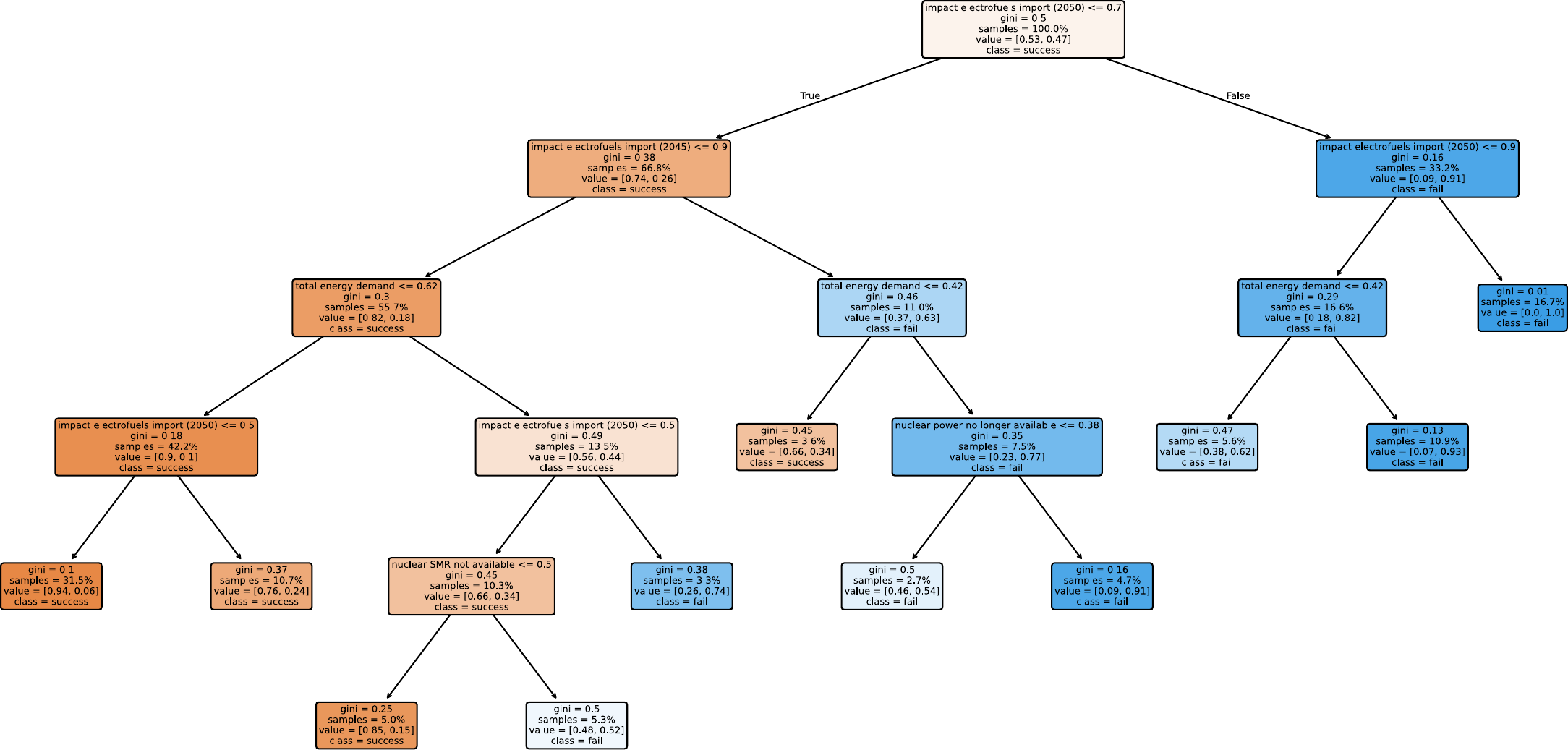}
\caption{
The classification tree depicting the primary branches leading to a successful or failed energy transition under perfect foresight for eleven leaves. The Gini index measures the impurity or disorder within the node, with lower values indicating purer (more homogeneous) nodes. ``samples" indicates the proportion of unexpected-event scenarios that end up in each node, while ``value" represent the share of those scenarios classified as success or failure. The class of each node corresponds to the label (success or failure) that is most prevalent within that node. Nodes are color-coded: blue for failure and orange for success. The color intensity reflects the proportion of scenarios, with deeper blue indicating a higher number of failed scenarios, and deeper orange indicating a higher number of successful scenarios. Related to Figure~2.
}
\label{fig:tree_true}
\end{figure}
\end{landscape}

\begin{figure}[h!]
\centering
\includegraphics[width=0.6\textwidth]{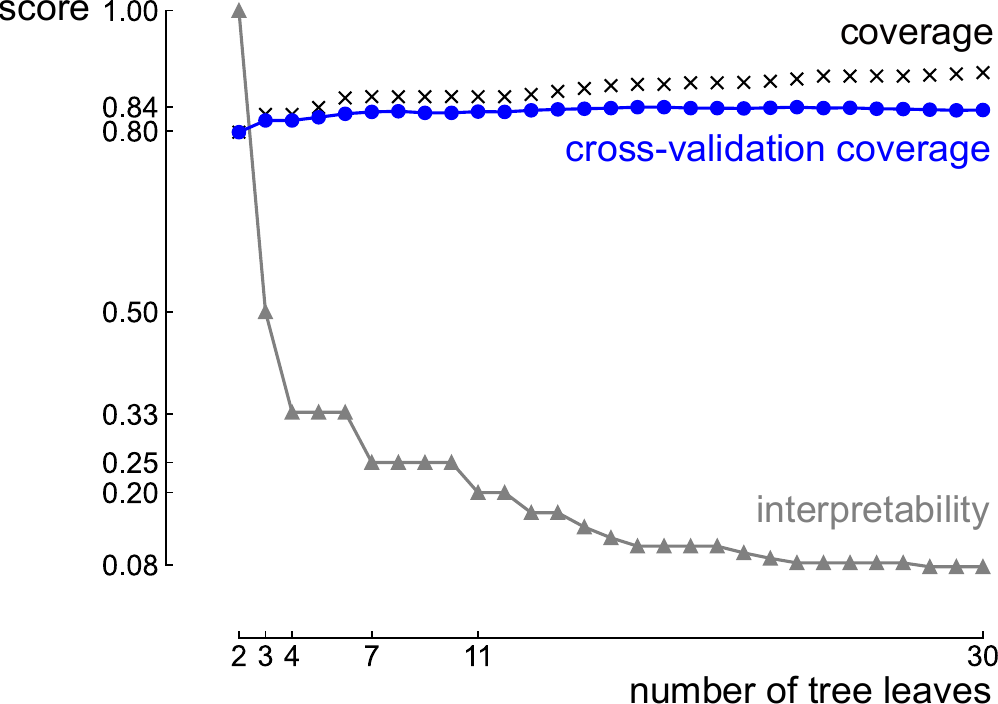}
\caption{The performance scores of the classification tree illustrates that a significant cross-validation coverage score (0.80) can be attained with two leaves and one unique identifier. The interpretability score is defined as the reciprocal of the unique number of features the classification tree uses for branching. A higher interpretability score indicates a simpler model, as fewer variables are involved. The coverage score, on the other hand, represents the proportion of data points correctly assigned to their respective clusters by the classification tree. Additionally, the coverage of a 5-fold cross-validation analysis is presented to show how consistently the model performs across different subsets of the data. Related to Figure~2.}
\label{fig:dt_sensitivity_n_leaves}
\end{figure}

The classification tree in the main paper was designed to highlight the key scenarios that lead to a failed energy transition (\autoref{fig:tree_true}). These scenarios are represented by final leaves with at least 90\% of the scenarios leading to failure. To determine the classification tree, we progressively increased the number of leaves and, at each step, measured interpretability, coverage, and cross-validation coverage using the method outlined by Baader~et~al.~\cite{baader2023streamlining}. The results indicate that cross-validation coverage reaches 0.80 with just two leaves and one identifier (\autoref{fig:dt_sensitivity_n_leaves}). In the final configuration with eleven leaves, the tree incorporates five unique identifiers and achieves the highest cross-validation coverage (0.84). Increasing the number of leaves further does not enhance coverage but decreases interpretability. 

\clearpage
\section{Convergence assessment of cumulative cost curves}
\label{sec:si:convergence_dt}

The accuracy of cumulative cost curves depends on the number of scenarios considered in their construction. To evaluate the adequacy of the number of unexpected-event scenarios used in assessing energy policy decisions, we analyzed two metrics and their convergence as more scenarios are considered.

\paragraph{Proportion of successful transitions} This metric examines the fraction of scenarios in which a successful energy transition is achieved. If we denote \( S \) as the total number of scenarios and \( S_{\text{success}} \) as the number of scenarios where the transition is successful, then the proportion of successful transitions is given by:
\begin{equation}
\text{Proportion}_{\text{success}} = \dfrac{S_{\text{success}}}{S}
\end{equation}

\paragraph{Area Under the Curve (AUC)} The area under the cumulative transition cost curve provides an aggregate measure of the transition costs across all scenarios. If \( C(x) \) represents the cumulative cost function, the area under the curve can be approximated by numerical integration methods. For discrete points \( x_1, x_2, \ldots, x_n \), the Area Under the Curve (AUC) is approximated by:
\begin{equation}
\text{AUC} \approx \sum_{i=2}^{n} \dfrac{C \left( x_{i} \right) + C \left(x_i-1 \right)}{2}  \left( x_{i+1} - x_i \right).
\end{equation}

\begin{figure}[h!]
\centering
\includegraphics[width=1\textwidth]{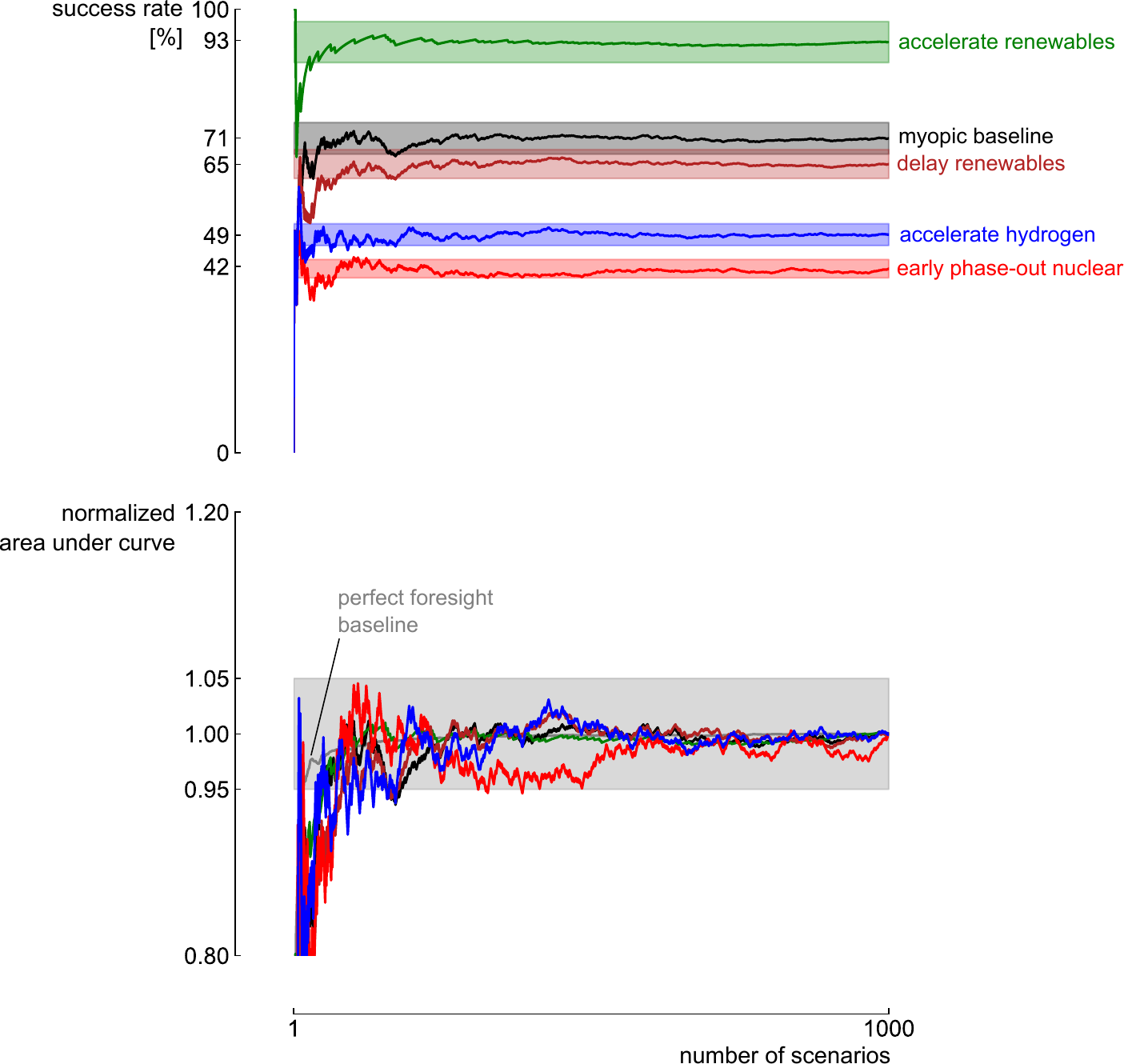}
\caption{The evolution of the convergence metrics with respect to the number of scenarios evaluated for the cumulative cost curves for each energy policy decision evaluated. The metrics stabilize within a 5\% bound after evaluating 1000 scenarios in the construction of the curves. The success rate of the perfect foresight curve is not shown, as it attains 100\% for the scenarios evaluated. Related to Figure~4.}
\label{fig:convergence}
\end{figure}

The evolution of these metrics with respect to the number of scenarios considered is shown in \autoref{fig:convergence}, which indicates sufficient accuracy is reached within 1000 scenarios.

\clearpage

\section{Diversity in investment decisions under perfect foresight}
\label{sec:si:designs_pf}

\begin{figure}[h!]
\centering
\includegraphics[width=1\textwidth]{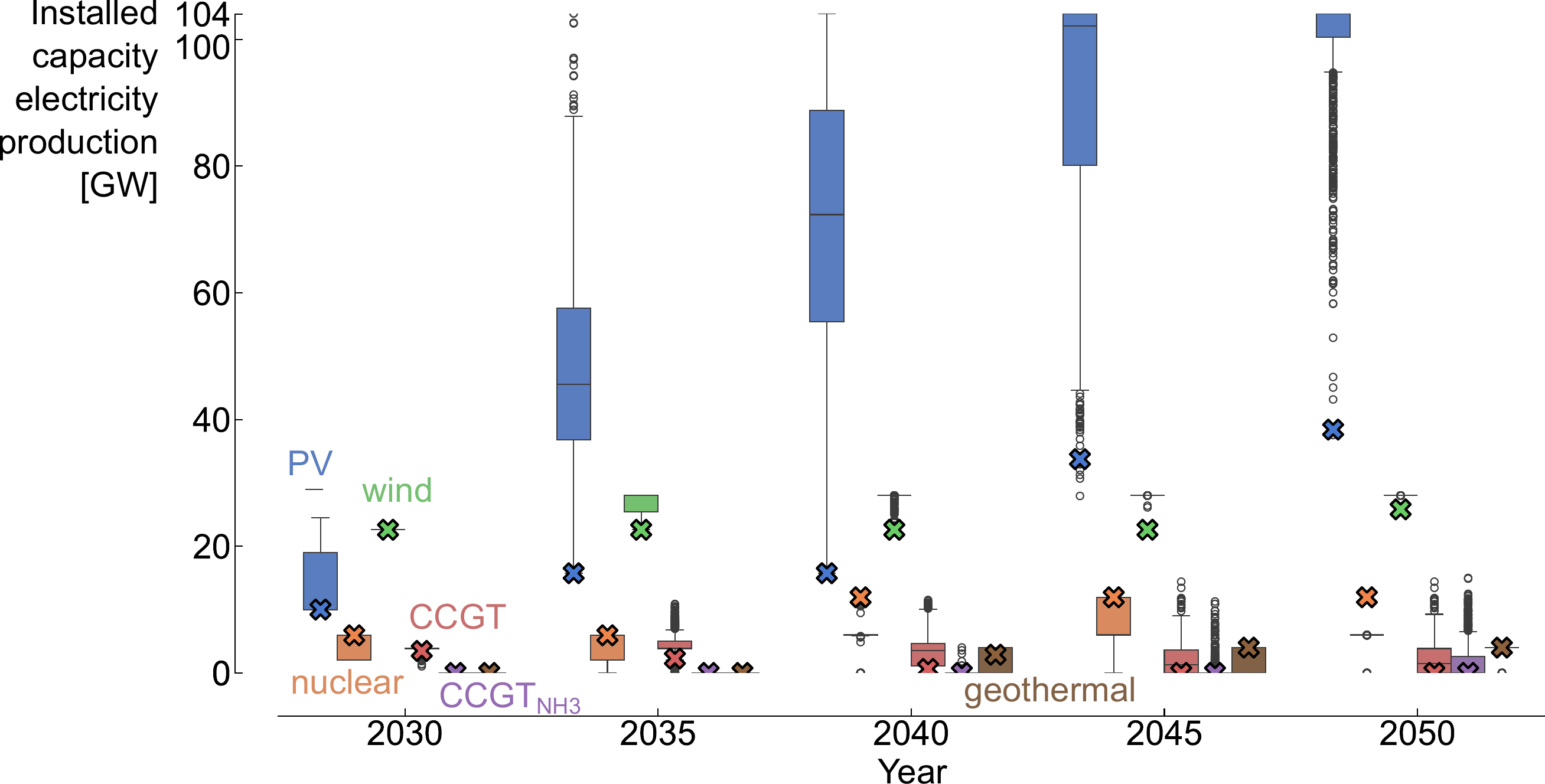}
\caption{The box plot illustrates the decisions made about installed capacities for power-producing technologies at each stage of the transition across various unexpected-event scenarios where a successful energy transition is possible under perfect foresight. The ``x'' marks the design choice in the absence of unexpected events. For power generation, the results show a shift from gas-powered to renewable technologies. In the baseline scenario (without unexpected events), the initial decision is to deploy the minimum capacity of PV and wind power, while maximizing nuclear power capacity, including both conventional and Small Modular Reactors (SMRs). However, under unexpected-event scenarios, PV and wind capacities are increased to compensate for disruptions to other energy supplies, such as the loss of electrofuel imports or nuclear power. Although ammonia-fired Combined Cycle Gas Turbines (CCGT) are not considered in the ideal scenario, they are frequently deployed under unexpected-event scenarios by 2050. By 2050, nuclear power may either be fully deployed (conventional + SMR), consist of only conventional nuclear power plants, or be absent due to exogenous impacts. Additionally, the deployment of geothermal power is restricted to either full deployment or none, depending on the unexpected event scenario. Related to Figure~2 in the main paper.}
\label{fig:freq_distr_electricity}
\end{figure}

\begin{figure}[h!]
\centering
\includegraphics[width=1\textwidth]{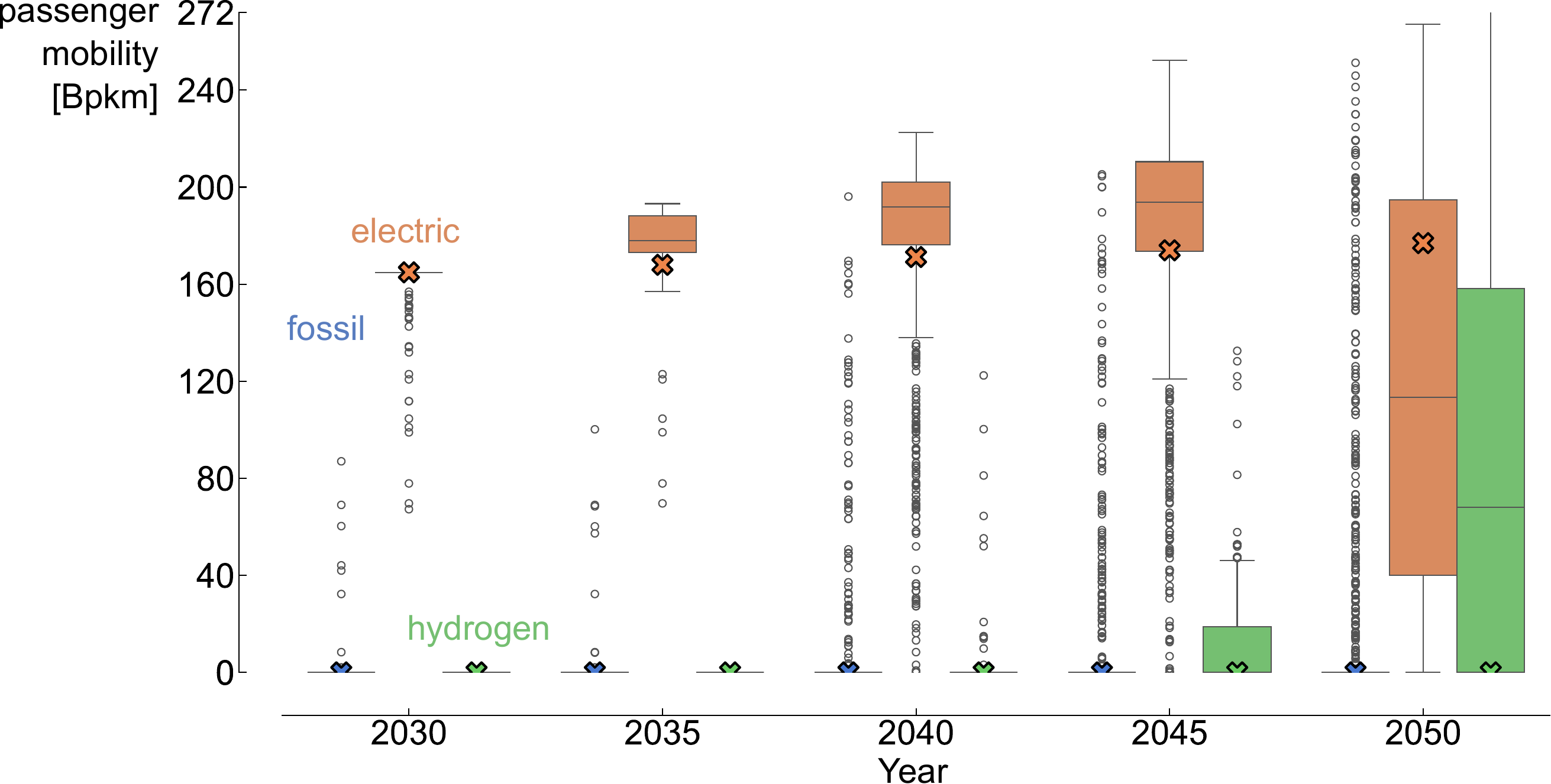}
\caption{The box plot illustrates the decisions made about installed capacities for passenger mobility technologies at each stage of the transition across various unexpected-event scenarios where a successful energy transition is possible under perfect foresight. The ``x'' marks the design choice in the absence of unexpected events. Decisions about passenger mobility indicate a rapid shift from fossil-fueled vehicles to electric vehicles, with only limited adoption of hydrogen-powered fuel cell vehicles until 2045. In 2050, when the electric vehicles deployed in 2030 reach the end of their life cycle, the model either renews the electric vehicle fleet---the preferred choice in an ideal future scenario---or switches to hydrogen-powered fuel cell vehicles, depending on the specific unexpected event scenario. Related to Figure~2 in the main paper.}
\label{fig:freq_distr_mobility}
\end{figure}

\begin{figure}[h!]
\centering
\includegraphics[width=1\textwidth]{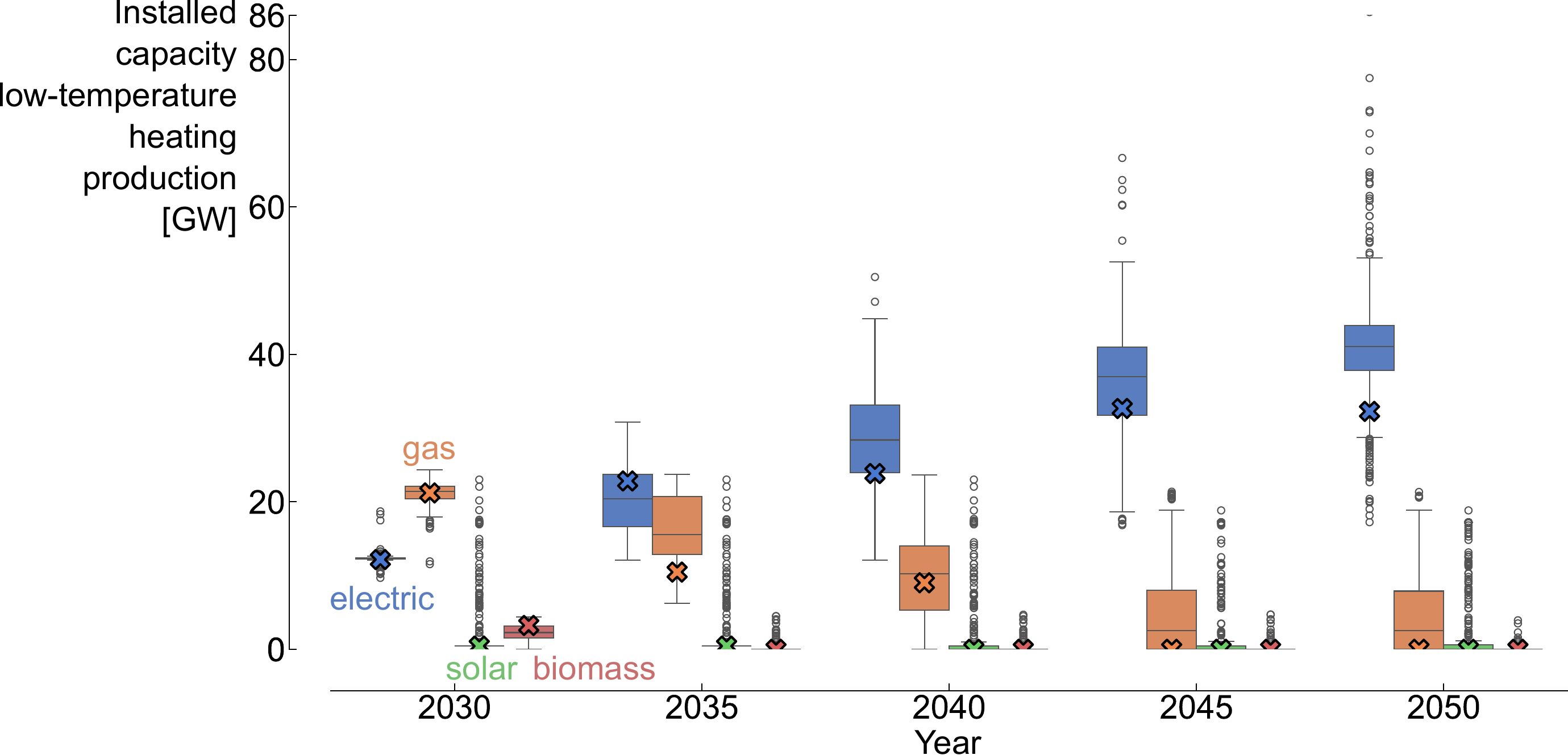}
\caption{The box plot illustrates the decisions made about installed capacities for low-temperature heating technologies, both decentralized and in District Heating Networks, at each stage of the transition across various unexpected-event scenarios where a successful energy transition is possible under perfect foresight. The ``x'' marks the design choice in the absence of unexpected events. There is a trend towards expanding electricity-powered technologies, such as heat pumps, while gradually phasing out gas-powered technologies throughout the transition (\autoref{fig:freq_distr_lt_heating}). Given Belgium's limited solar potential, solar-powered technologies are seldom considered in this context. Although biomass-fired technologies (wood, hydrolysis, wet biomass) are of interest in 2035, their adoption is rare in subsequent years. Related to Figure~2 in the main paper.}
\label{fig:freq_distr_lt_heating}
\end{figure}

\begin{figure}[h!]
\centering
\includegraphics[width=1\textwidth]{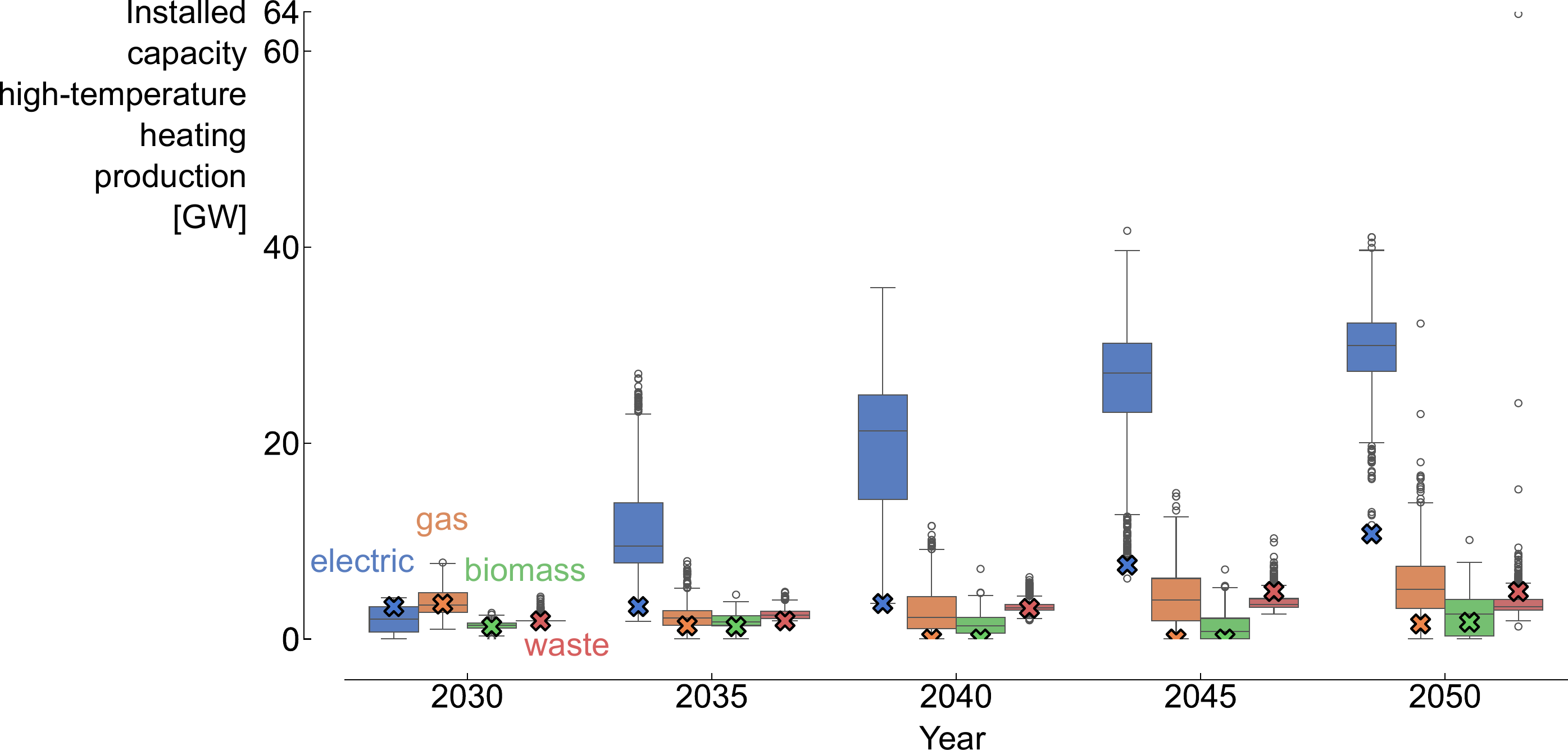}
\caption{The box plot illustrates the decisions made about installed capacities for high-temperature heating technologies at each stage of the transition across various unexpected-event scenarios where a successful energy transition is possible under perfect foresight. The ``x'' marks the design choice in the absence of unexpected events. Investment decisions for industrial high-temperature heating reveal a diverse mix of technologies, including electricity-powered, gas-fired, biomass-fired, and municipal solid waste-fired options. In an ideal scenario without unexpected events, there is a slightly higher investment in electric-powered technologies. However, this represents the minimum capacity for electric heating when unexpected events occur, suggesting that to manage such scenarios, more high-temperature electric heating is required. Related to Figure~2 in the main paper.}
\label{fig:freq_distr_ht_heating}
\end{figure}

\begin{figure}[h!]
\centering
\includegraphics[width=1\textwidth]{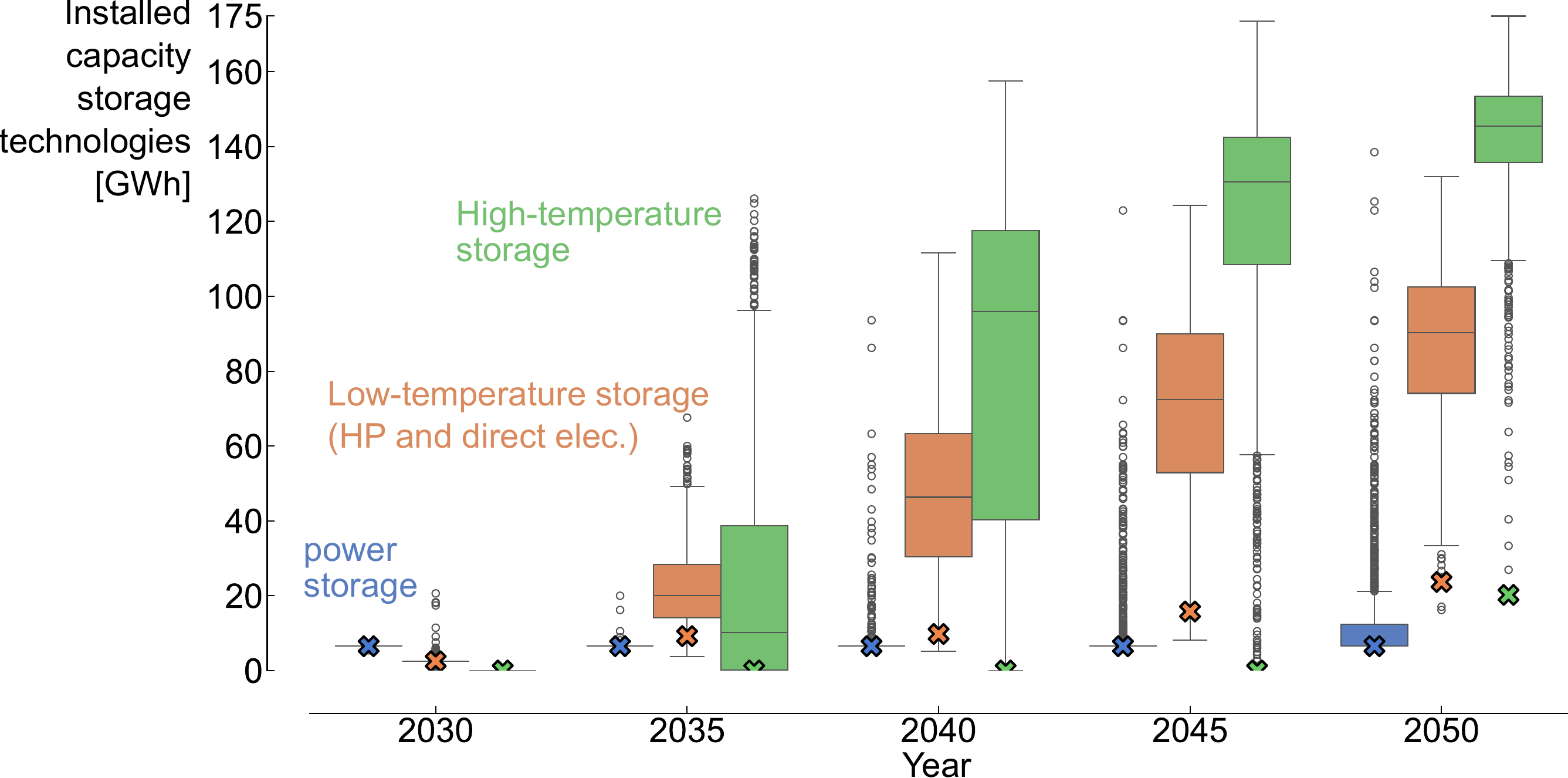}
\caption{The box plot illustrates the decisions made about installed capacities for energy storage technologies at each stage of the transition across various unexpected-event scenarios where a successful energy transition is possible.  The results indicate a growing emphasis on high-temperature storage over time. The ``x'' on the plot marks the design choice in scenarios without unexpected events, where minimal storage deployments are planned. The decisions regarding storage reveal a limited use of electric storage technologies (such as pumped hydro storage and Li-ion batteries), as well as low-temperature and high-temperature thermal storage in the ideal scenario. However, with the increased deployment of intermittent renewable power (\autoref{fig:freq_distr_electricity}) and electric heating (\autoref{fig:freq_distr_lt_heating} and \autoref{fig:freq_distr_ht_heating}) in unexpected-event scenarios, the capacities of storage technologies are also increased in those cases. Related to Figure~2 in the main paper.}
\label{fig:freq_distr_storage}
\end{figure}

\begin{figure}[h!]
\centering
\includegraphics[width=1\textwidth]{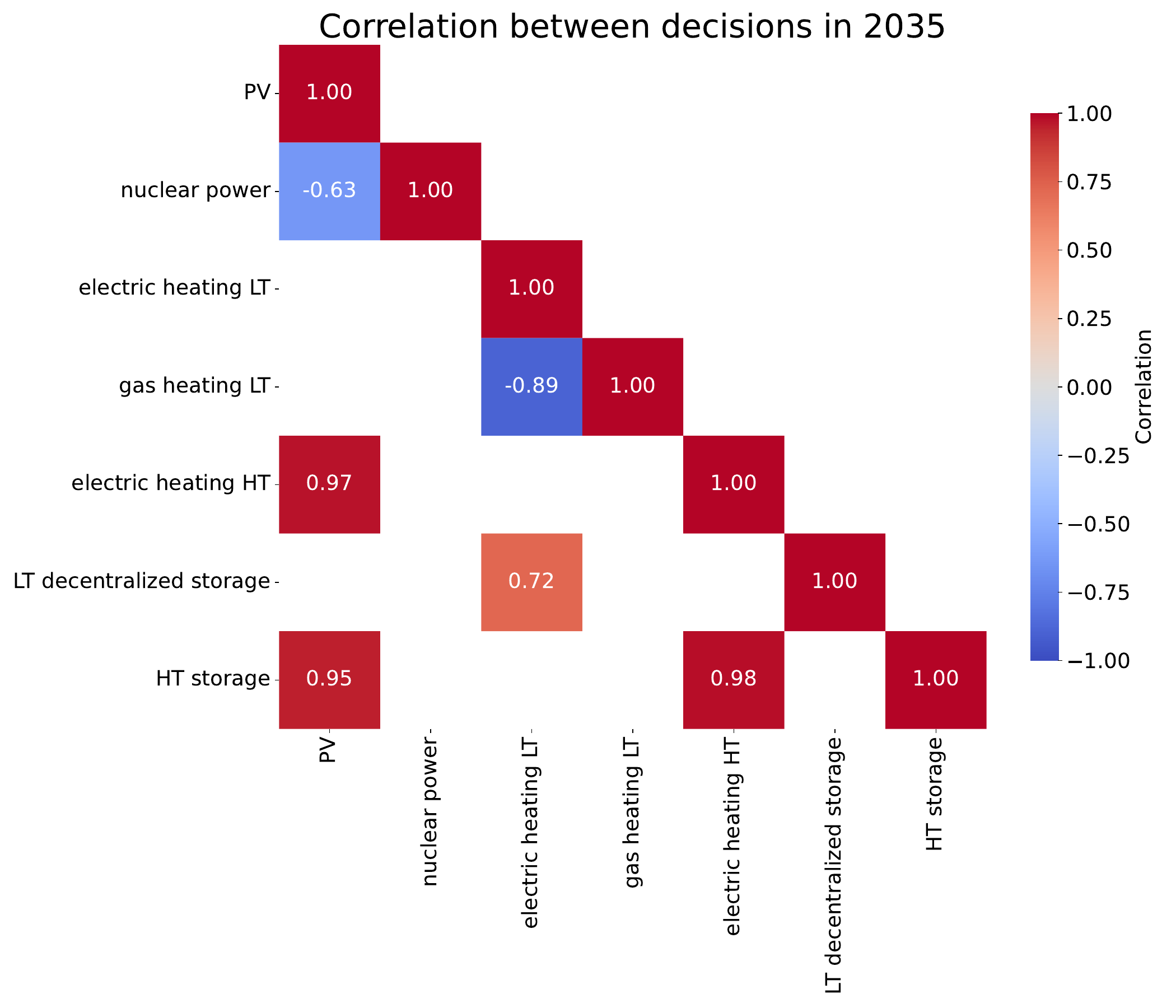}
\caption{The correlation matrix for investment decisions in 2035 under perfect foresight reveals a strong positive correlation between the deployment of PV and high-temperature electric heating, suggesting that increased investment in PV is closely associated with a rise in high-temperature electric heating capacity. Conversely, PV deployment negatively correlates with nuclear power, indicating that higher investments in PV are necessary when nuclear power is no longer available. Additionally, gas heating and electric low-temperature heating show a strong negative correlation, reflecting a preference for one over the other in investment strategies. Correlations with an absolute value below 0.6 are masked in the matrix to focus on more significant relationships. Investment decisions that do not exhibit a correlation with any other technology above this threshold are excluded from the matrix. Related to Figure~2 in the main paper.}
\label{fig:corr_2035}
\end{figure}

\begin{figure}[h!]
\centering
\includegraphics[width=1\textwidth]{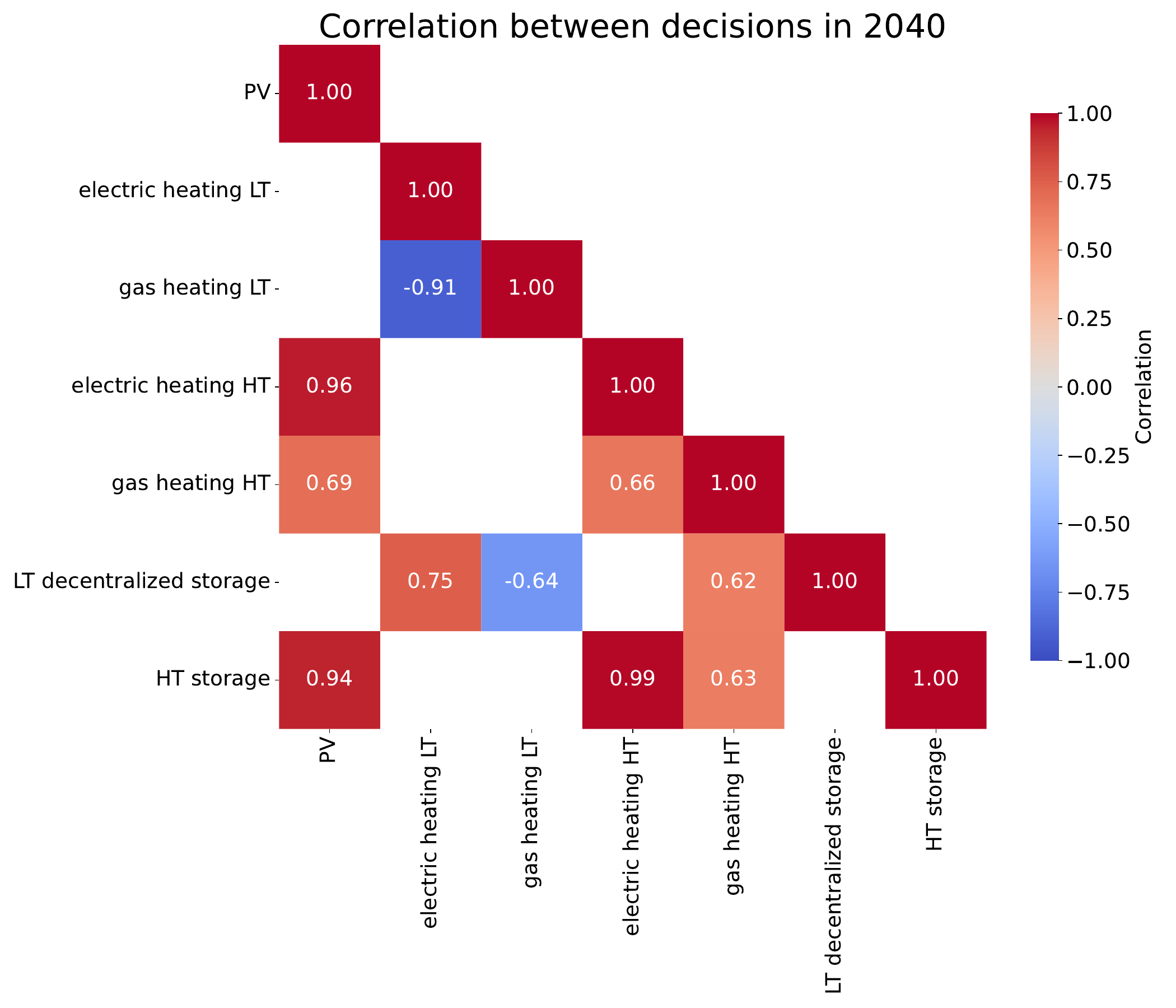}
\caption{The correlation matrix for investment decisions in 2040 under perfect foresight reveals a strong positive correlation between the deployment of PV and High-Temperature (HT) electric heating, suggesting that increased investment in PV is closely associated with a rise in high-temperature electric heating capacity. Additionally, gas heating and electric Low-Temperature (LT) heating show a strong negative correlation, reflecting a preference for one over the other in investment strategies. HT storage is positively correlated with both electric and gas HT heating, as it accommodates heat storage regardless of the source. Meanwhile, LT decentralized storage, which supports heat from electric-powered LT heaters, is positively correlated with LT electric heating and negatively correlated with gas LT heating. Correlations with an absolute value below 0.6 are masked in the matrix to focus on more significant relationships. Investment decisions that do not exhibit a correlation with any other technology above this threshold are excluded from the matrix. Related to Figure~2 in the main paper.}
\label{fig:corr_2040}
\end{figure}

\begin{figure}[h!]
\centering
\includegraphics[width=1\textwidth]{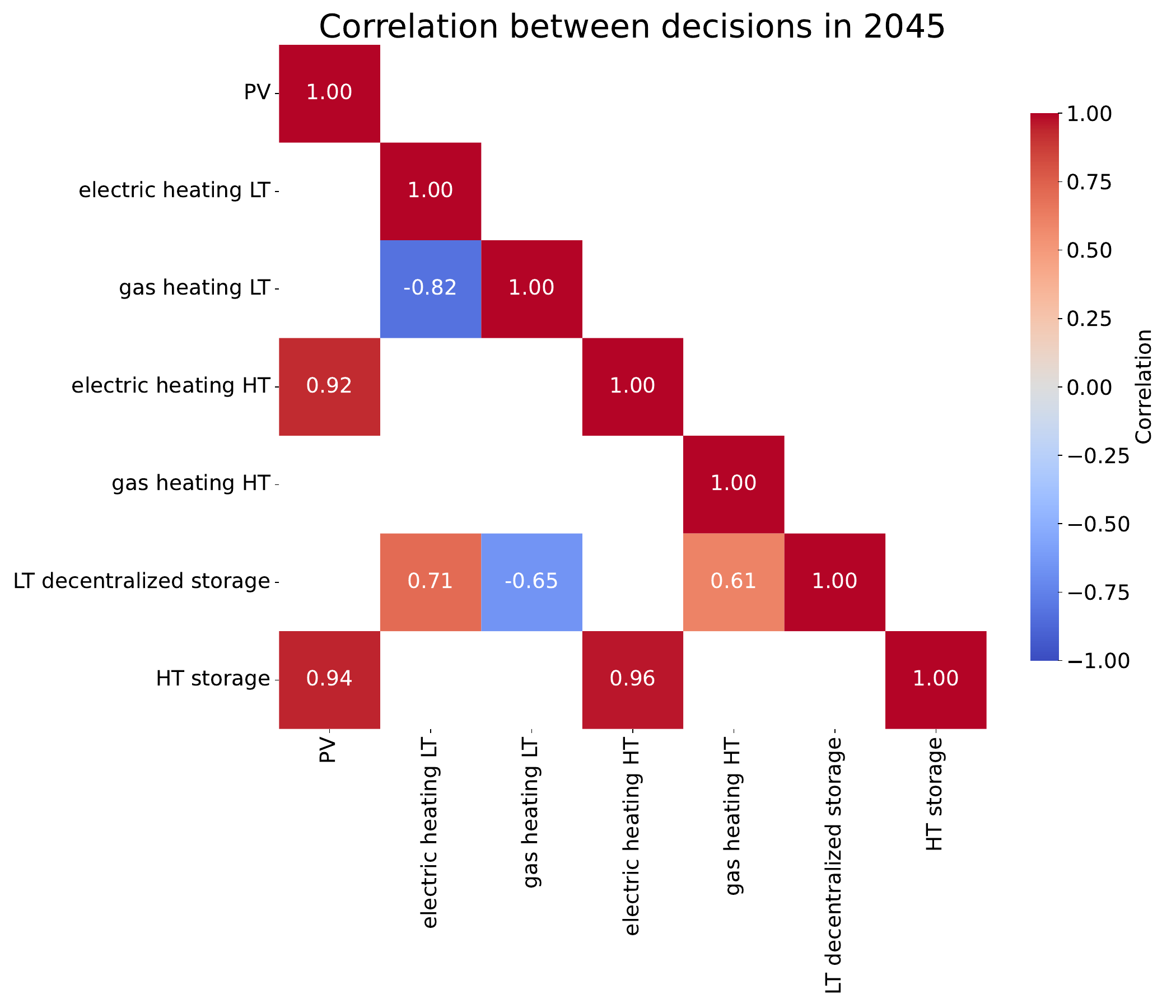}
\caption{The correlation matrix for investment decisions in 2045 under perfect foresight reveals a strong positive correlation exists between the deployment of PV and high-temperature electric heating, suggesting that increased investment in PV is closely associated with a rise in high-temperature electric heating capacity. Additionally, gas heating and electric low-temperature heating show a strong negative correlation, reflecting a preference for one over the other in investment strategies. Correlations with an absolute value below 0.6 are masked in the matrix to focus on more significant relationships. Investment decisions that do not exhibit a correlation with any other technology above this threshold are excluded from the matrix. Related to Figure~2 in the main paper.}
\label{fig:corr_2045}
\end{figure}

\begin{figure}[h!]
\centering
\includegraphics[width=1\textwidth]{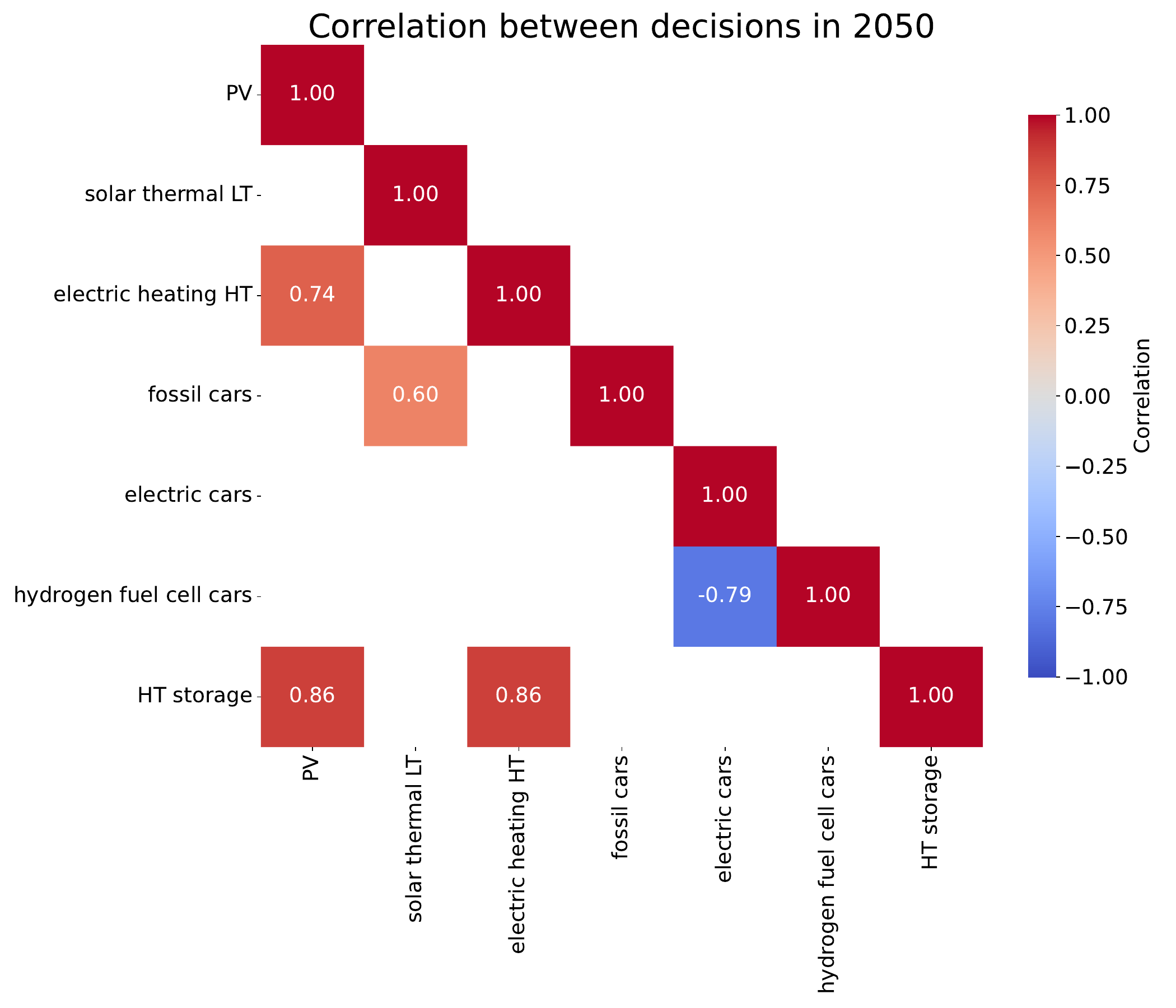}
\caption{The correlation matrix for investment decisions in 2050 under perfect foresight reveals that PV, high-temperature electric heating, and high-temperature storage technologies are closely correlated. In contrast, hydrogen-powered fuel cell vehicles are negatively correlated with electric vehicles, reflecting divergent investment strategies that prioritize one form of transportation over the other. To focus on significant relationships, correlations with an absolute value below 0.6 are masked in the matrix. Any investment decision without a correlation above this threshold with other technologies is removed. Related to Figure~2 in the main paper.}
\label{fig:corr_2050}
\end{figure}

\begin{figure}[h!]
\centering
\includegraphics[width=1\textwidth]{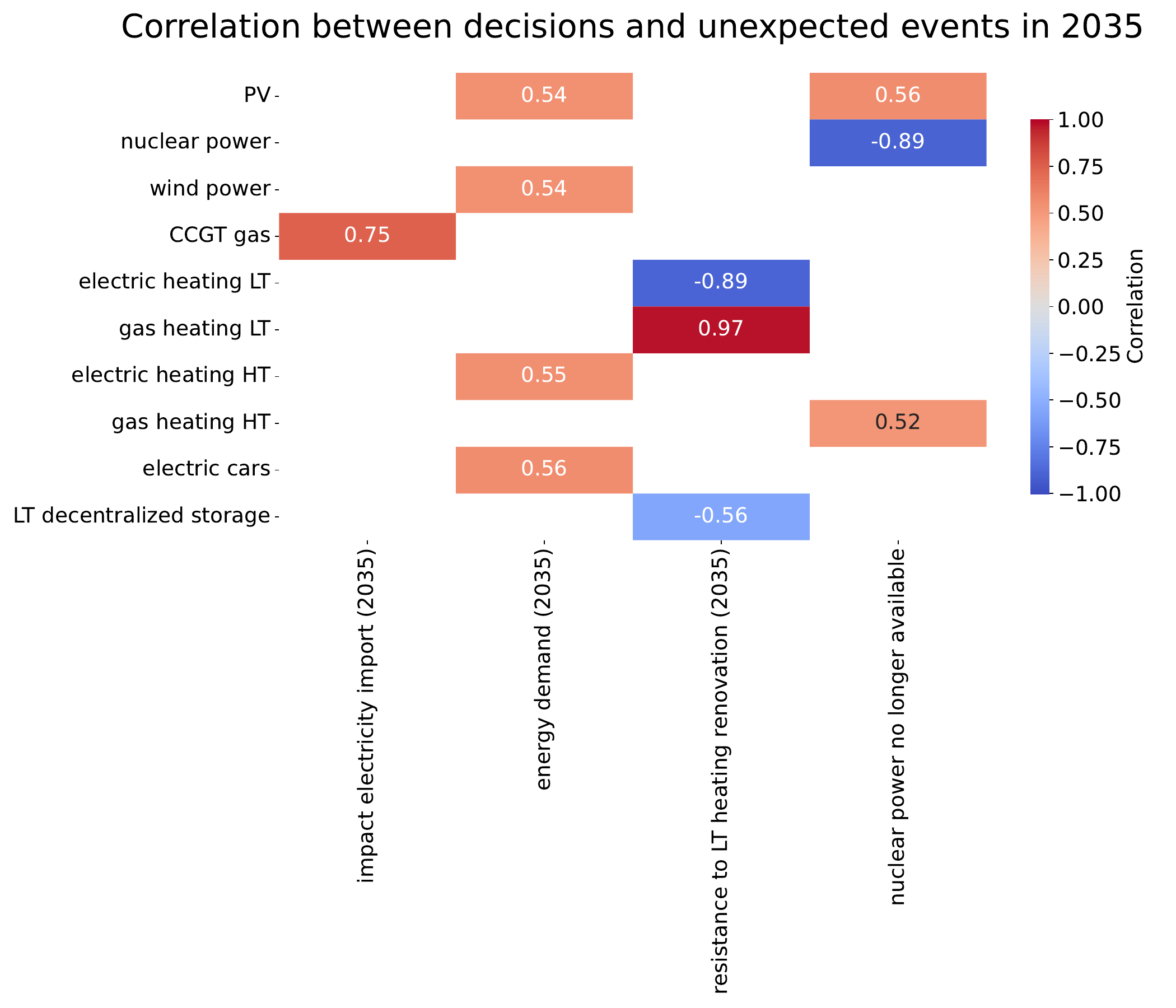}
\caption{The correlation between decisions and unexpected event impacts in 2035 shows that when electricity imports are affected, investments in gas-fired CCGTs increase to compensate for the reduced electricity supply. When final energy demand rises, it is offset by increased investments in PV and wind power, as well as High-Temperature (HT) electric heating and electric cars. There is also a clear positive correlation between resistance to Low-Temperature (LT) heating renovation and the size of existing LT gas-fired heating technology; the greater the resistance, the larger these systems remain. Conversely, there is a strong negative correlation between electric-powered LT heating and resistance to LT heating renovation. Finally, nuclear power capacity is negatively correlated with the impact of nuclear power becoming unavailable---the greater the impact, the sooner it is phased out, which is offset by increased investments in PV. Correlations between decisions and parameters subject to unexpected events with an absolute value below 0.5 are masked in the matrix to focus on more significant relationships. Investment decisions that do not exhibit a correlation with any parameter subject to unexpected events above this threshold are excluded from the matrix. Related to Figure~2 in the main paper.}
\label{fig:corr_decision_impacts_2035}
\end{figure}

\begin{figure}[h!]
\centering
\includegraphics[width=1\textwidth]{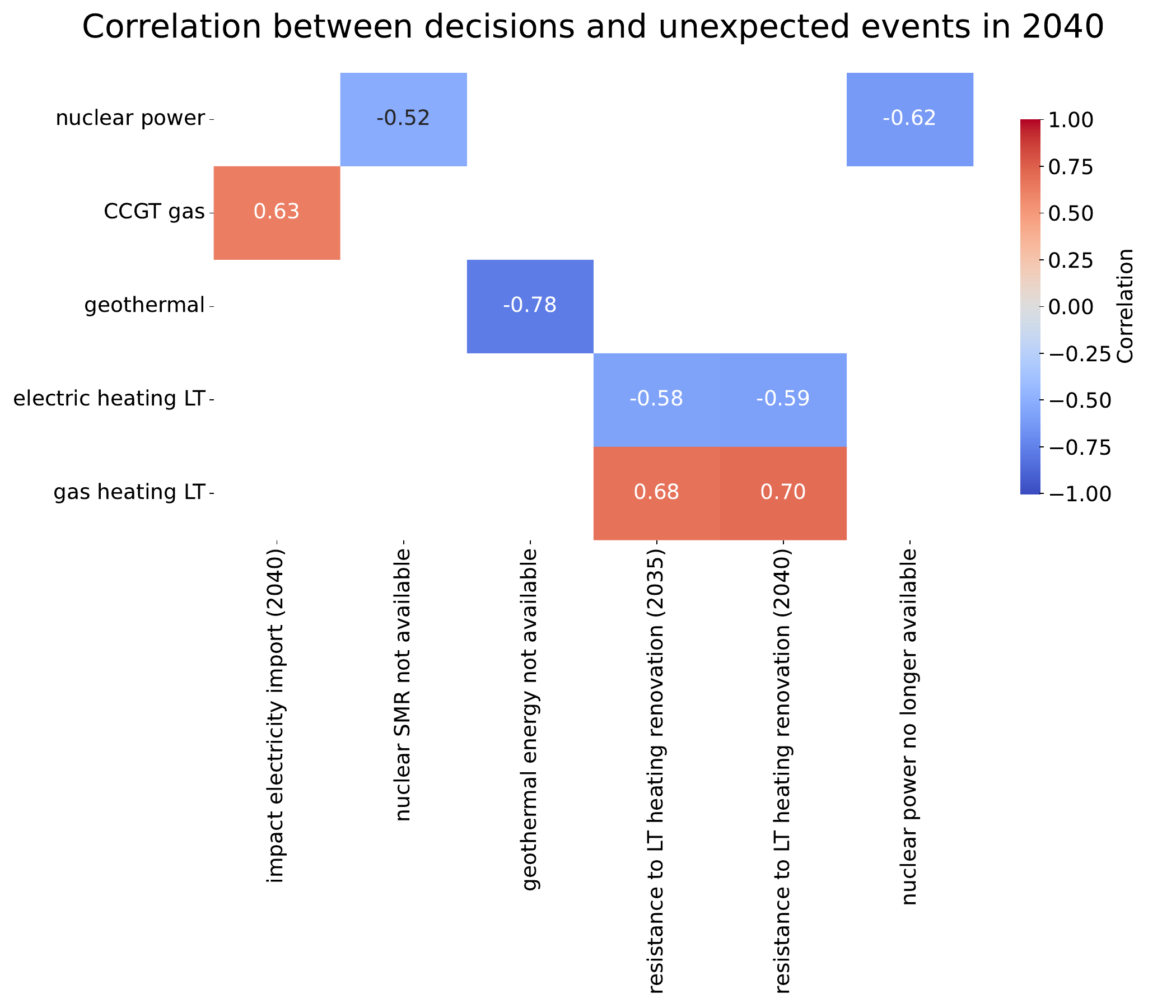}
\caption{The correlation between decisions and unexpected event impacts in 2040 shows that when electricity imports are affected, investments in gas-fired CCGTs increase to compensate for the reduced electricity supply. There is a clear positive correlation between resistance to Low-Temperature (LT) heating renovation and the size of existing LT gas-fired heating technology; the greater the resistance, the larger these systems remain. Similarly, there is a strong negative correlation between electric-powered LT heating and resistance to LT heating renovation. Finally, nuclear power capacity (both conventional and Small Modular Reactors) is negatively correlated with the impact of nuclear power becoming unavailable---the greater the impact, the sooner it is phased out---and nuclear SMR being not available in 2040. Correlations between decisions and parameters subject to unexpected events with an absolute value below 0.5 are masked in the matrix to focus on more significant relationships. Investment decisions that do not exhibit a correlation with any parameter subject to unexpected events above this threshold are excluded from the matrix. Related to Figure~2 in the main paper.}
\label{fig:corr_decision_impacts_2040}
\end{figure}

\begin{figure}[h!]
\centering
\includegraphics[width=1\textwidth]{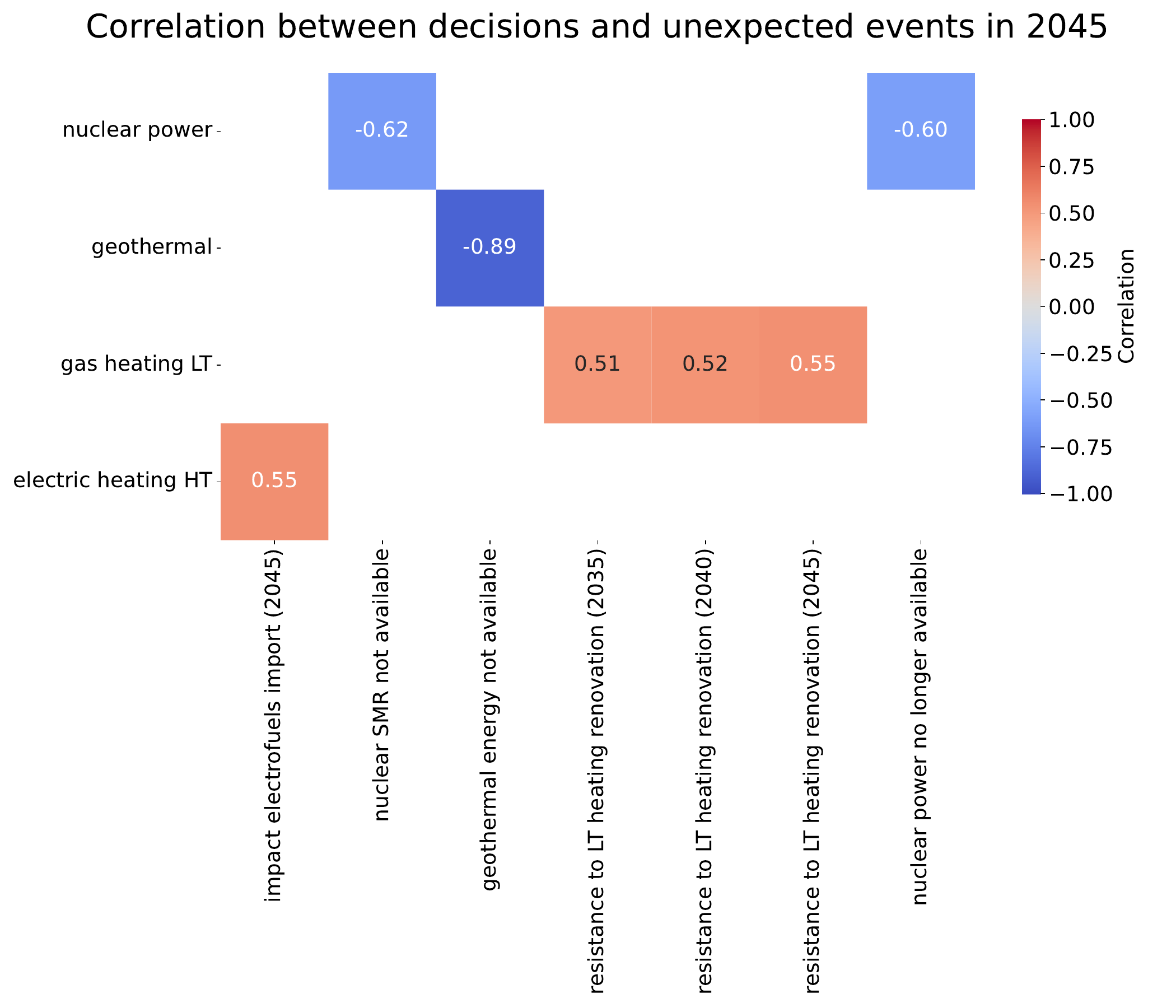}
\caption{The correlation between decisions and unexpected event impacts in 2045 shows that a positive correlation exists between the loss of electrofuel imports and increased capacity in PV power to compensate low-carbon energy supply losses, which is then used for high-temperature electric heating. There is also a positive correlation between resistance to Low-Temperature (LT) heating renovation and the size of existing LT gas-fired heating technology. Finally, nuclear power capacity (both conventional and Small Modular Reactors, or SMRs) is negatively correlated with the impact of nuclear power becoming unavailable---the greater the impact, the sooner it is phased out---and nuclear SMRs are not available in 2045. Correlations between decisions and parameters subject to unexpected events with an absolute value below 0.5 are masked in the matrix to focus on more significant relationships. Investment decisions that do not exhibit a correlation with any parameter subject to unexpected events above this threshold are excluded from the matrix. Related to Figure~2 in the main paper.}
\label{fig:corr_decision_impacts_2045}
\end{figure}

\begin{figure}[h!]
\centering
\includegraphics[width=1\textwidth]{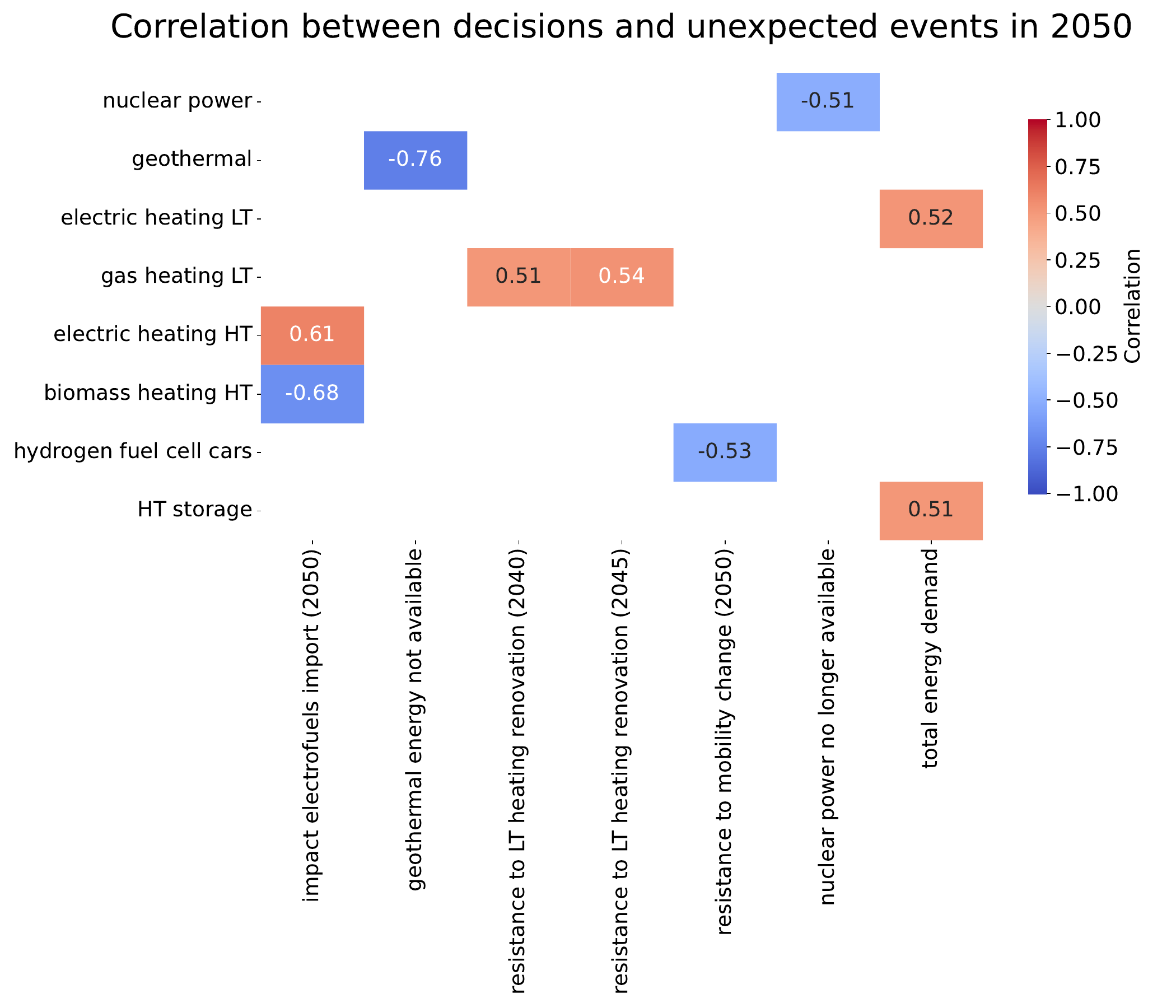}
\caption{The correlation between decisions and unexpected event impacts in 2050 shows that there is a positive correlation between the loss of electrofuel imports and increased electrification of High-Temperature (HT) heating. Additionally, there is a positive correlation between resistance to Low-Temperature (LT) heating renovation and the size of existing LT gas-fired heating technology. Interestingly, there is a negative correlation between resistance to mobility change and the ability to switch from electric to hydrogen fuel cell cars, indicating that the transition to hydrogen-powered passenger mobility is feasible only if resistance at that time is limited. Finally, nuclear power capacity (both conventional and Small Modular Reactors, or SMRs) is negatively correlated with the impact of nuclear power becoming unavailable—the greater the impact, the sooner it is phased out—and nuclear SMRs are not available in 2050. Correlations between decisions and parameters subject to unexpected events with an absolute value below 0.5 are masked in the matrix to focus on more significant relationships. Investment decisions that do not exhibit a correlation with any parameter subject to unexpected events above this threshold are excluded from the matrix. Related to Figure~2 in the main paper.}
\label{fig:corr_decision_impacts_2050}
\end{figure}

In 53\% of the unexpected-event scenarios, the perfect foresight approach succeeds in achieving an energy transition that meets climate targets at reasonable cost. However, each scenario results in different investment decisions regarding which technologies to adopt and when to expand them or decommission others. This section illustrates the wide range of investment decisions made across the evaluated unexpected-event scenarios under perfect foresight in power generation (\autoref{fig:freq_distr_electricity}), mobility (\autoref{fig:freq_distr_mobility}), low-temperature heating (\autoref{fig:freq_distr_lt_heating}), high-temperature heating (\autoref{fig:freq_distr_ht_heating}), and storage (\autoref{fig:freq_distr_storage}). Given the diversity of these decisions across the decision space, we also present the correlation between all decisions in 2035 (\autoref{fig:corr_2035}), 2040 (\autoref{fig:corr_2040}), 2045 (\autoref{fig:corr_2045}), and 2050 (\autoref{fig:corr_2050}). Finally, we show the correlation between the impacts on the parameters subject to unexpected events and the decisions made in 2035 (\autoref{fig:corr_decision_impacts_2035}), 2040 (\autoref{fig:corr_decision_impacts_2040}), 2045 (\autoref{fig:corr_decision_impacts_2045}), and 2050 (\autoref{fig:corr_decision_impacts_2050}).

\clearpage

\clearpage

\clearpage

\clearpage

\clearpage

\clearpage

\clearpage

\clearpage

\clearpage

\clearpage

\clearpage

\clearpage

\clearpage

\clearpage

\section{Early-stage energy policy decisions}
\label{sec:si:energypolicydecisions}

We introduce several illustrative energy policy decisions into the model to demonstrate the functionality of the method. For each policy, we forced several investment decisions in 2030 and allowed the model to optimize the remaining decisions in 2030 and the pathway beyond 2030 using a myopic approach. 

The first policy decision is based on Belgium's Recovery and Resilience Plan under the REPowerEU chapter~\cite{beresplan}, which aims at accelerating the energy transition by prioritizing PV and wind deployment, easing administrative barriers, and promoting electrification in heating and mobility. Funding will be directed toward heat pumps in both the residential and the commercial sectors, supported by infrastructure upgrades to handle increased electricity demand. To represent the potential of this policy, we enforce the full deployment of PV, onshore and offshore wind power capacity by 2030 as an exogenous input, while allowing other technology capacities to remain flexible. When optimizing the pathway myopically, the early availability of abundant renewable electricity results in an expanded heat pump distribution in centralized and decentralized low-temperature heating, increases the role of direct electrification in high-temperature industrial heating, and supports the shift to an electrified mobility fleet.

In contrast, we also evaluate a decision characterized by further delaying the transition. While this is not a current energy policiy, recent years have seen delays due to political short-termism, public misunderstanding, and competing economic interests~\cite{stern2015we}. Therefore, it is relevant to explore the consequences of continued inaction in the context of unexpected events. The second evaluated decision---delaying the transition---restricts a capacity expansion for PV, onshore and offshore wind between 2025 and 2030.

The third decision focuses on the early phase-out of nuclear power, based on the nuclear phase-out law~\cite{nuclearphaseoutlaw}. To represent its impact, nuclear power (including SMR) was removed from the technology options starting in 2030.

The fourth and final decision involves making directed upfront investments in hydrogen-powered technologies identified as necessary or promising levers for decarbonization by the Federal Hydrogen Strategy~\cite{fedh2strat}. This includes the exogenous imposition of 150 MW of electrolyzers by 2030. Fuel cell heavy-duty trucks and methanol-powered boats were deployed to meet heavy-duty mobility demands, while fuel cell buses are deployed to cover expected bus mobility needs. Additionally, to represent heating in buildings where heat pumps and district heating are challenging, 10\% of the low-temperature heating demand is met by hydrogen-powered cogeneration units.

\begin{figure}[h!]
\centering
\hspace{-4cm}
\includegraphics[width=0.7\textwidth]{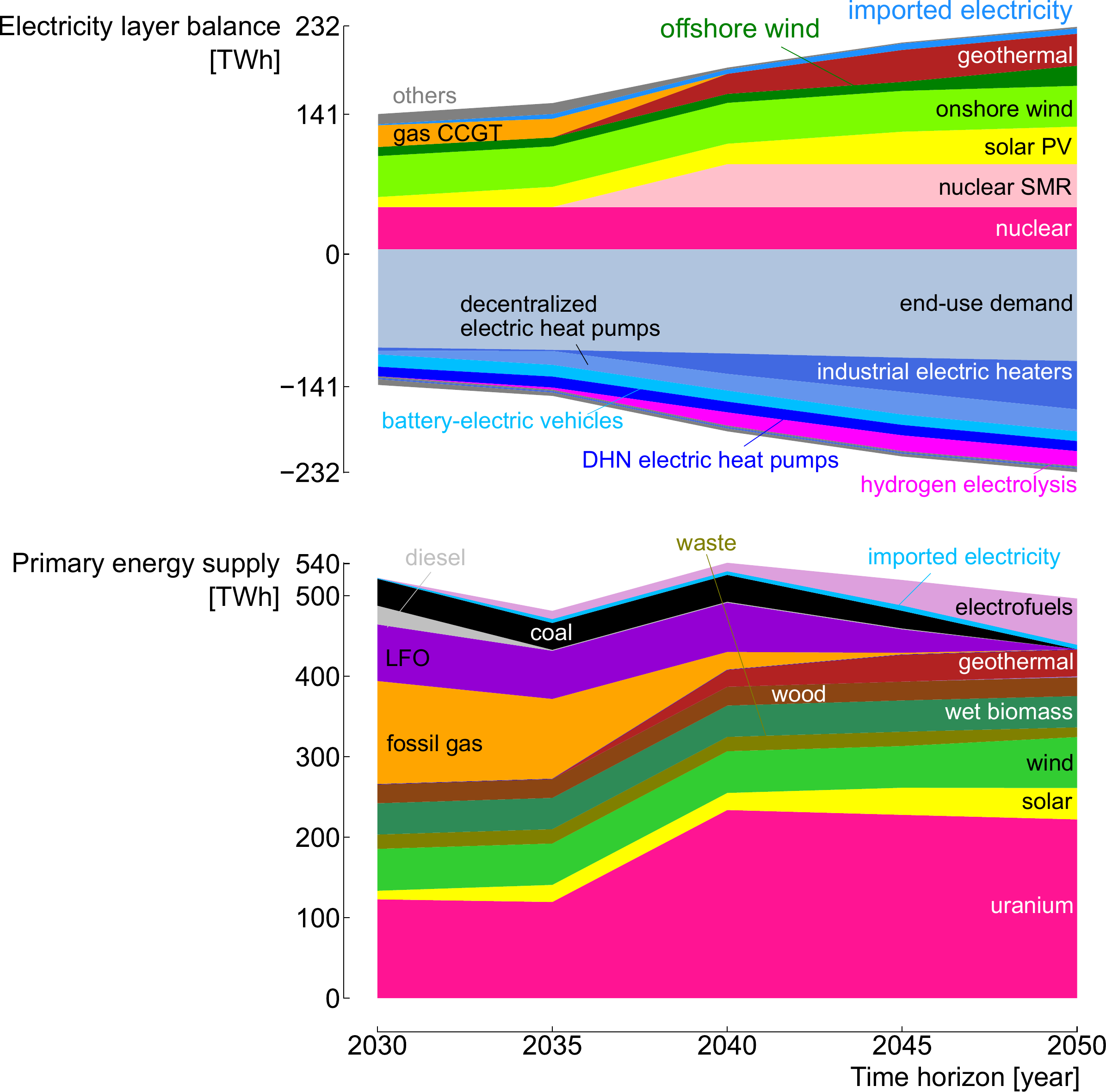}
\caption{For the myopic baseline decision, the electricity layer balance and primary energy supply reveal a shift in energy sources over time. Uranium initially plays a pivotal role, powering both conventional nuclear plants and, from 2040 onward, small modular reactors (SMRs). By 2030, onshore wind energy reaches its peak capacity, while photovoltaic (PV) and offshore wind capacities begin to rise steadily, enriching the electricity mix. The introduction of geothermal power in 2040 adds further diversity to the energy portfolio. As these renewable sources and new technologies are integrated, a gradual phase-out of fossil fuels is set in motion. Fossil gas, light fuel oil (LFO), diesel, and coal gradually make way for cleaner alternatives. By 2045, electrofuels become a significant import, and local hydrogen production through electrolysis accelerates. On the consumption side, heat pumps and battery-electric vehicles rapidly meet their target capacities, while industrial electric heaters increase steadily over time. Related to Figure~4.}
\label{fig:layers_resources_baseline}
\end{figure}

These energy policy decisions were compared against a myopic baseline, which represents the myopic pathway without imposing any early-stage decisions. If no unexpected events occur, this pathway unfolds as follows: By 2030, onshore wind energy reaches peak capacity, while PV and offshore wind capacities increase steadily (\autoref{fig:layers_resources_baseline}). The introduction of geothermal power and nuclear SMR in 2040 further diversifies the energy portfolio. As renewable sources and new technologies are integrated, fossil fuels are gradually phased out: fossil gas, light fuel oil (LFO), diesel, and coal are progressively replaced by cleaner alternatives. By 2045, electrofuels are significantly imported, and local hydrogen production through electrolysis accelerates. On the consumption side, heat pumps and battery-electric vehicles reach their target capacities quickly, while industrial electric heaters increase steadily.

\begin{figure}[h!]
\centering
\includegraphics[width=1\textwidth]{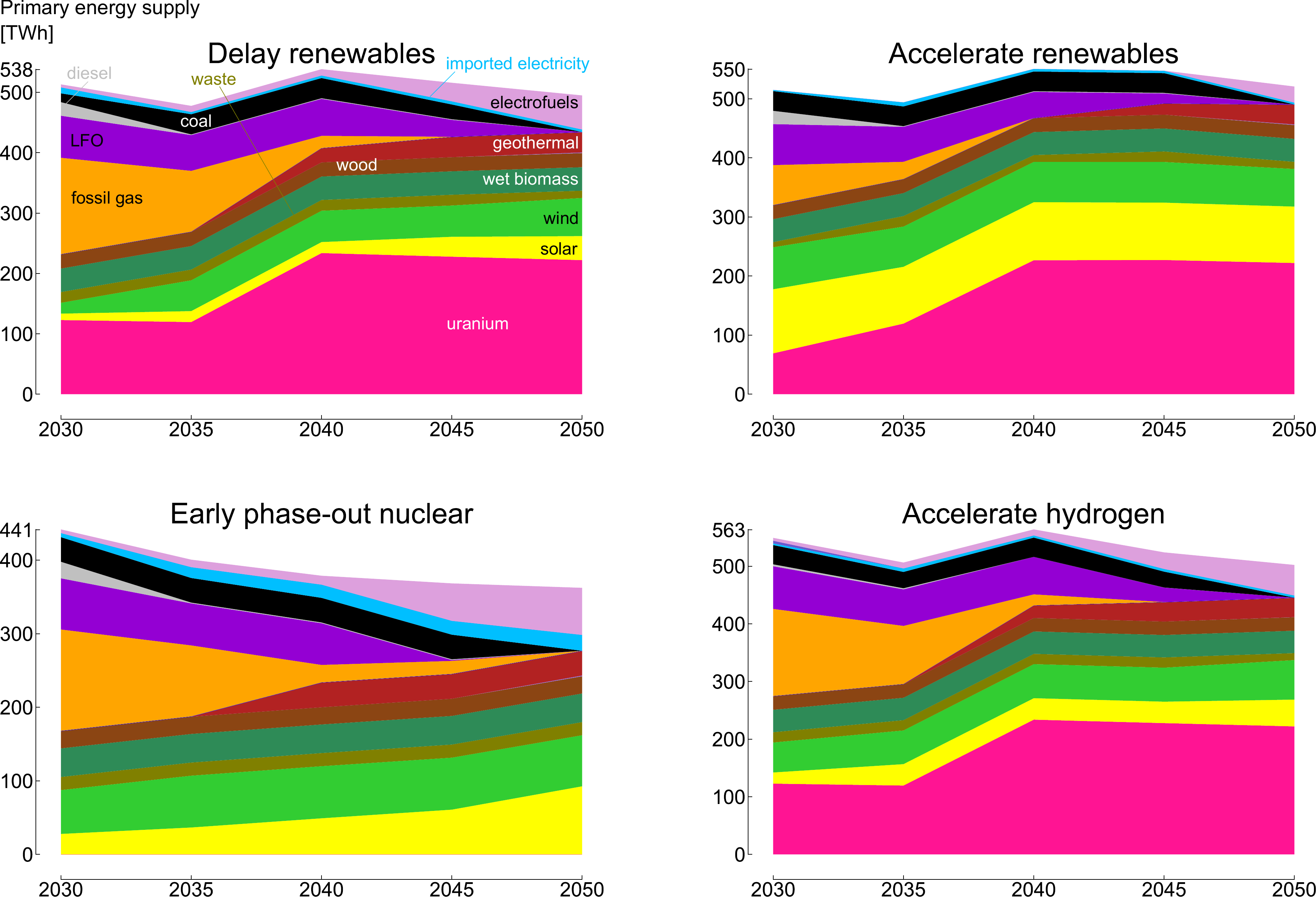}
\caption{The primary energy supply highlights the effects of various transition strategies. Delaying renewable deployment prolongs dependence on fossil gas, whereas accelerating renewables significantly reduces fossil gas imports by increasing wind and solar capacity. However, an early nuclear phase-out requires a rapid scale-up of PV capacity, delaying the reduction of fossil gas use and increasing reliance on electricity imports and electrofuels later in the transition. Similarly, accelerating hydrogen production slightly raises electrofuel imports and increases dependence on solar energy. Related to Figure~4 in the main paper.}
\label{fig:resources_1}
\end{figure}

\begin{figure}[h!]
\centering
\includegraphics[width=1\textwidth]{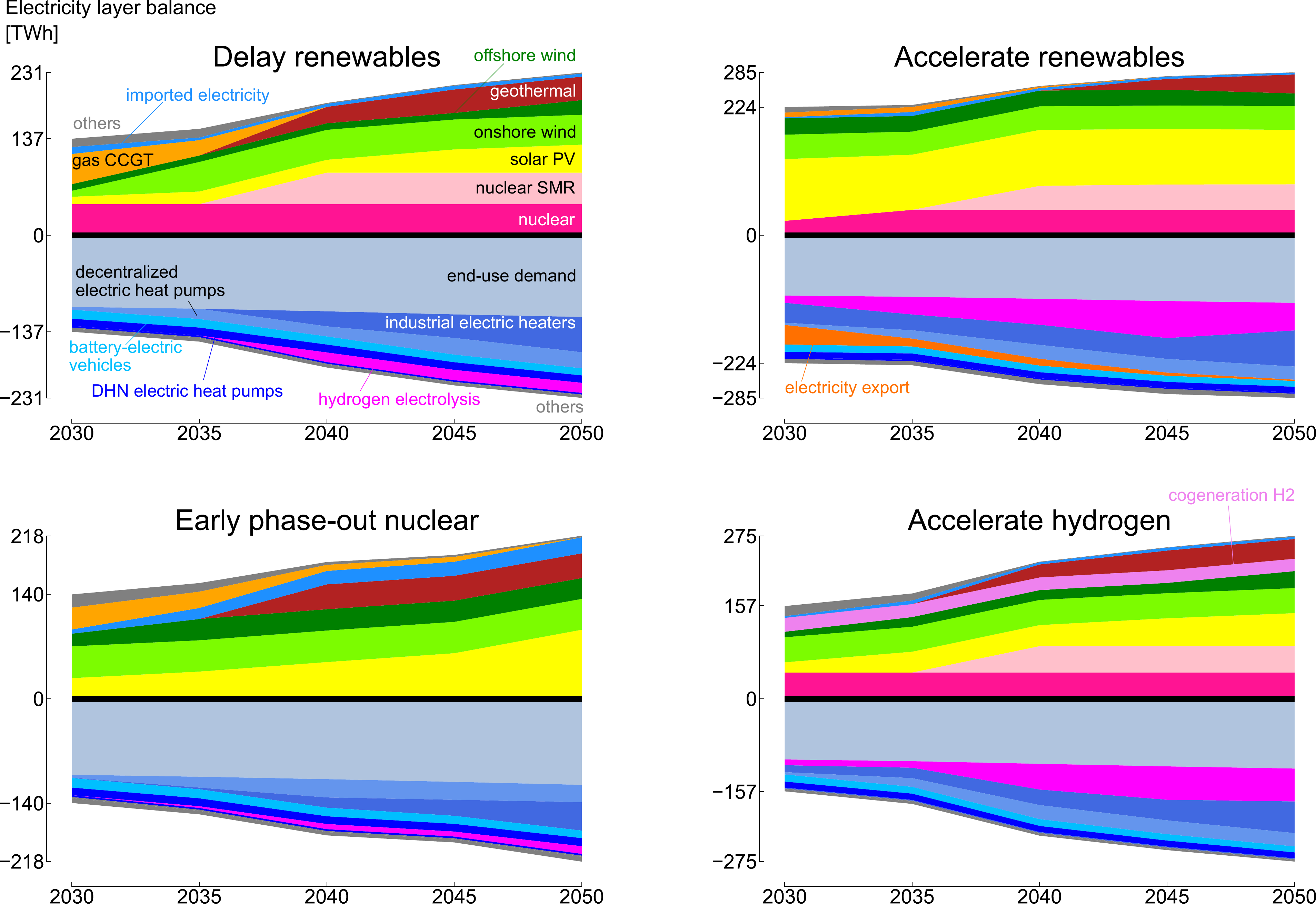}
\caption{The electricity layer balances reveal the impact of different transition strategies. Delaying the deployment of renewables prolongs reliance on gas-powered combined cycle gas turbines (CCGTs). On the other hand, accelerating the deployment of renewables leads to a rapid increase in photovoltaic (PV) and wind energy, resulting in a slower upgrade of nuclear power capacity, significant local hydrogen production through electrolysis and electricity exports. Moreover, due to the extensive renewable power installed, it does not rely on a full deployment of geothermal power later in the transition. In contrast, an early phase-out of nuclear power significantly affects the electricity layer balance, necessitating an increased reliance on photovoltaic (PV) capacity from the start. This shift also slows down the phase-out of gas-powered combined CCGTs, increases dependence on electricity imports, and results in minimal local hydrogen production. Accelerating hydrogen focuses on quickly ramping up local hydrogen production---reaching an electricity balance of 275~TWh by 2050---to cover the hydrogen-powered cogeneration units. Related to Figure~4 in the main paper. }
\label{fig:layers_1}
\end{figure}

Similarly, the primary energy supply and electricity layer balance for the pathways under expected conditions starting from the different early-stage decisions are presented in \autoref{fig:resources_1} and \autoref{fig:layers_1}, respectively. Delaying renewable deployment extends reliance on fossil gas (\autoref{fig:resources_1}, top left) and gas-powered CCGTs (\autoref{fig:layers_1}, top left). Conversely, accelerating renewable deployment leads to a rapid increase in PV and wind energy, slower upgrades of nuclear power, significant local hydrogen production through electrolysis, and electricity exports in 2030 (\autoref{fig:layers_1}, top right). This extensive renewable power reduces the need for geothermal power later. An early phase-out of nuclear power significantly affects the electricity layer, requiring more PV capacity from the start (\autoref{fig:resources_1} and \autoref{fig:layers_1}, bottom left). This shift slows the phase-out of gas-powered CCGTs, increases dependence on electricity imports, and results in minimal local hydrogen production. Accelerating hydrogen focuses on quickly ramping up local hydrogen production---reaching an electricity balance of 275~TWh by 2050---to cover the hydrogen-powered cogeneration units (\autoref{fig:layers_1}, top right). 

\end{appendices}

\end{document}